\documentclass[%
 reprint,
 amsmath,amssymb,
 aps,
]{revtex4-2}
\usepackage{xcolor}
\usepackage{graphicx}
\usepackage{dcolumn}
\usepackage{bm}
\usepackage{amsmath}
\DeclareMathOperator{\sgn}{sgn}

\begin{document}

\preprint{APS/123-QED}

\title{ Magneto-optical transport in type-II Weyl semimetals in the presence of orbital magnetic moment}


\author{Panchlal Prabhat}
\affiliation{ Department of Physics, Lalit Narayan Mithila University, Darbhanga, Bihar 846004, India}%

\author{Amit Gupta}
\email{sunnyamit31@gmail.com}
\affiliation{
  Department of Physics, M. R. M. College, Lalit Narayan Mithila University, Darbhanga, Bihar 846004, India}



\date{\today}

\begin{abstract}
In this work, we investigate the linear and nonlinear magnetoconductivity of tilted type-II Weyl semimetals within the semiclassical Boltzmann framework by incorporating both the Berry curvature (BC) and the orbital magnetic moment (OMM) corrections. In the linear response regime, we derive analytical expressions for the magnetoconductivity that explicitly include the previously ignored OMM contribution while in the nonlinear regime we obtain analytical expressions that account for the combined effects of the BC and OMM. We find that the open Fermi surface and finite density of states characteristic of the type-II phase qualitatively alter the BC and OMM-induced transport responses relative to the type-I case. Depending on the magnetic-field orientation and transport geometry, the OMM is shown to either suppress or enhance different magnetoconductivity components, in contrast to their monotonic behavior in type-I Weyl semimetals. Our results provide a comprehensive theoretical framework for magneto-optical transport in tilted type-II Weyl semimetals and identify experimentally accessible signatures of the interplay between cone tilt, BC and OMM in candidate materials such as WTe$_2$, MoTe$_2$ and TaIrTe$_4$.
\end{abstract}

\maketitle


\section{\label{sec:level1}
Introduction
 }
In 1929, Hermann Weyl hypothesized a novel form of fermion with zero mass and explicit chirality, known as the Weyl fermion (WF) \cite{weyl1929electron}. Weyl semimetals(WSMs) are 3D topological materials that host this WF and have been experimentally realized in condensed matter systems\cite{xu2015discovery,lv2015experimental,lu2015experimental, kim2013dirac, huang2015weyl}. In the Brillouin zone, these materials manifest themselves as a pair of conduction and valence bands that touch at a discrete location known as the Weyl nodes(points)\cite{murakami2007phase, wan2011topological, yang2011quantum, burkov2011weyl, xu2011chern, huang2015weyl, lv2015observation}. The Nielsen–Ninomiya theorem requires Weyl nodes to occur in pairs of opposite chirality within the Brillouin zone \cite{nielsen1981no}. Each Weyl node acts as a monopole source or sink of Berry curvature (BC), which is typically regarded as an effective magnetic field in momentum space. This geometric structure underlies several anomalous transport phenomena, including the anomalous Hall effect \cite{berry1984quantal,xiao2010berry}, negative magnetoresistance associated with the chiral anomaly \cite{huang2015observation} and surface Fermi arcs \cite{potter2014quantum,moll2016transport}.\\

The BC structure of Weyl bands gives rise to a broad range of magnetotransport and thermoelectric phenomena. Semiclassical treatments of magnetotransport in Weyl systems have been developed by incorporating BC corrections into the carrier dynamics \cite{mccormick2017semiclassical}. Chiral-anomaly-related transport and the associated anomalous responses have been investigated in Refs. \cite{nielsen1983adler,zyuzin2012topological,son2012berry,son2012chiral}. The semiclassical description of the anomalous Hall effect and BC effects in metals was established in Refs. \cite{sinitsyn2007semiclassical,haldane2004berry}, while anomalous transport of Weyl fermions was further studied in Ref. \cite{landsteiner2014anomalous}. Thermoelectric responses of Weyl systems have also been investigated theoretically \cite{ferreiros2017anomalous,laurell2015thermoelectric,chen2016thrmoelectric}. These studies provide the theoretical foundation for the semiclassical magneto-optical transport considered in the present work.\\

The WFs in WSMs are not restricted by Lorentz symmetry, in contrast to their high-energy counterpart, which permits the Weyl cones to tilt in energy-momentum space. When the tilting of the cones is large enough such that the velocity changes sign along the tilting direction, a transition from type-I to type-II WSMs occurs \cite{soluyanov2015type,li2017evidence,autes2016robust,jiang2017signature,pal2018optical,
zhang2018coexistence,chang2016prediction,xu2017discovery}. When the tilt velocity remains smaller than the Fermi velocity, the material is classified as a type-I Weyl semimetal, where the Fermi surface encloses isolated electron or hole pockets. However, once the tilt exceeds the Fermi velocity, the Weyl cone becomes overtilted and undergoes a Lifshitz transition into the type-II phase. In this regime, electron and hole pockets coexist at the Fermi energy, producing an open Fermi surface and a finite density of states even at the Weyl node \cite{soluyanov2015type}. These unique electronic structures strongly modify anomalous thermoelectric and Hall responses  \cite{saha2018anomalous,zyuzin2016intrinsic,mukherjee2018anomalous}, chiral-anomaly-induced longitudinal magnetotransport \cite{sharma2017chiral} and distinguish type-II Weyl semimetals from their type-I counterparts.\\

Besides BC, another intrinsic geometric property of Bloch electrons is the
orbital magnetic moment (OMM) $\bm{m}_{\bm{k}}^s$, which arises from the self-rotation of the electronic wave packet around its center of mass \cite{sundaram1999wave}. In the presence of an external magnetic field, the OMM modifies the quasiparticle energy according to
\begin{equation}
\tilde{\epsilon_{\bm k}^s} = \epsilon_{\bm k}^s -\bm{m}_{\bm{k}}^s\cdot \bm{B}
\end{equation}
This energy correction subsequently affects the group velocity, the equilibrium distribution function and the nonequilibrium transport response. Its role is conceptually distinct from the direct anomalous-velocity contribution of the BC and from the BC correction to the phase-space density of states. Thus, although the OMM and BC are intimately connected geometric properties of Bloch bands, they enter the semiclassical transport theory through different physical mechanisms. In the present work, the magnetic-field dependence of the conductivity is therefore modified not by introducing an additional magnetic degree of freedom, but by incorporating the intrinsic magnetic moment of the Bloch wave packet into the semiclassical dynamics. Previous theoretical studies have shown that the OMM plays an important role in the magneto-optical response of untilted and type-I WSMs, where it generally suppresses the magnetoconductivity through partial cancellation of the BC contribution \cite{morimoto2016semiclassical, gao2022suppression}.\\

Although magnetotransport in type-II Weyl semimetals has been investigated from several
perspectives, including linear magnetoresistance, planar Hall transport, nonlinear optical
responses and Landau-level quantization, the combined influence of strong cone tilting
and the OMM on magneto-optical transport has received comparatively
little attention \cite{das2019linear, ghosh2024direction}.

The present work addresses this problem by employing a systematic semiclassical
Boltzmann theory of linear and nonlinear magneto-optical transport in tilted type-II WSMs, including both BC and OMM corrections. In the linear response regime, we derive analytical expressions for the magnetoconductivity that explicitly include the previously ignored OMM contribution, while in the nonlinear regime we obtain analytical expressions that account for the combined effects of the BC and OMM.  Analytical expressions
are obtained for the longitudinal magnetoconductivity, planar Hall conductivity, ordinary
Hall conductivity, quadratic magnetoconductivity and second-harmonic generation under
different magnetic-field configurations. Our analysis demonstrates that the OMM produces
substantially stronger corrections in several conductivity components in the type-II phase than in type-I WSMs because of the modified Fermi-surface topology. We further show that all conductivity components explicitly depend on the ultraviolet momentum cutoff, reflecting the noncompact Fermi surface of the low-energy continuum model rather than a universal property of Weyl semimetals.\\


Magnetic Weyl semimetals (MWSMs) constitute another important class of topological materials in which Weyl nodes emerge from the breaking of time-reversal symmetry by intrinsic magnetic order. In contrast to their nonmagnetic counterparts, MWSMs can exhibit large intrinsic BC, anomalous Hall conductivity, and pronounced magneto-optical responses even in the absence of an external magnetic field \cite{shi2018prediction, li2019intrinsic, liu2019magnetic, wang2019single, soh2019ideal, ma2019spin, nie2022magnetic, sun2023magnetic,  bose2025anomalous}. These properties make them particularly suitable for investigating both linear and nonlinear magneto-optical transport(e.g. Berry curvature dipole(BCD) induced nonlinear Hall effects, second-harmonic generation) with magnetic order providing an additional control parameter for tuning the electronic and optical responses \cite{sun2023magnetic}.\\

The remainder of the paper is organized as follows. In Sec.\ref{model_ham}, we introduce the tilted Weyl Hamiltonian and summarize the semiclassical Boltzmann formalism including BC and OMM corrections. Sections  \ref{linear_response} and \ref{non-linear} present the analytical derivation of the linear and nonlinear magneto-optical conductivities under different magnetic-field configurations. We end with conclusions in Sec. \ref{conclusion}.\\

\section{Model Hamiltonian and Semiclassical Boltzmann approach}\label{model_ham}

The effective non-interacting low-energy Hamiltonian for tilted WSMs is given by \cite{zyuzin2016intrinsic},
\begin{eqnarray}
\label{eq:ham}
\mathcal{H}^s_0=\hbar \mathit{v}_F (s\bm{k}\cdot\bm{\sigma} + t_s k_z\sigma_{0})
\end{eqnarray}

\noindent where $s={\pm}1$ is the chirality of the Weyl node, $\bm{k}$ is the momentum, $\mathit{v}_F$ is the Fermi velocity,  $\bm{\sigma}{\{\sigma_x, \sigma_y,\sigma_z}\}$  are the three Pauli matrices and $\sigma_{0}$ is a $2\times2$ identity matrix, the parameter $t_s$ characterizes the tilting of the Weyl cone. The energy dispersion for tilted WSMs is given by $ \epsilon_{\bm{ k}}^s=\hbar \mathit{v}_F (t_sk_z{\pm}{k})$ with ${\pm}$ representing the conduction and valence bands respectively. This Hamiltonian corresponds to type-I WSM only for a small tilt parameter  $t_s\textless 1$, such that the Fermi surface encloses only an electron $(\mu\textgreater 0)$ or a hole $(\mu\textless 0)$ pocket. And for type-II WSM, the tilt parameter  $t_s\textgreater 1$, leading over tilted Weyl nodes and at the Fermi energy, the presence of unbounded electron and hole pockets. Our results for conductivity elements for type-II WSMs are written irrespective of the sign of the tilt parameter $t_s$. This helps us to understand the even and odd nature of the conductivity elements of the system.\\

We will use semiclassical Boltzmann equations in this study. In the presence of a static magnetic field $ \bm{B}$ and a time varying electric field $\bm {E}$, the semiclassical equations of motion at the location $\textbf{r}$ and the wave-vector  $\textbf{k}$ in a given band are \cite{sundaram1999wave,xiao2010berry}
\begin{eqnarray}
\dot{\bm r}=\bm{\tilde{v}}_{ \bm{k}}^s-\dot{\bm{k}}\times \Omega_{\bm{k}}^s\label{EOMa}\\
\hbar \dot{\bm k}=-e\bm{E}-e\dot{\bm{r}}\times \bm{B}\label{EOMb}
\end{eqnarray}

\noindent where -e is the electron charge. The first term on the right-hand side of Eq.(\ref{EOMa}) is defined as group velocity with $\bm{\tilde{v}}_{ \bm{k}}^s = \frac{1}{\hbar} \nabla_{\bm k} \tilde{\varepsilon}_{\bm{ k}}^s$ =$\bm{v}_{ \bm{k}}^s-\frac{1}{\hbar}
\nabla_{\mathbf k}(\bm{m}_{\bm{k}}^s\cdot \bm{B})$,  $\bm{v}_{\bm{k}}^s =\frac{1}{\hbar} \nabla_{\bm k} \varepsilon_{\bm{ k}}^s$. The BC and OMM are defined as \cite{sundaram1999wave,xiao2010berry}

\begin{equation}
{\bm \Omega}_{\bm k}^s= Im[\langle \nabla_{\bm k} u_{\bm k}^s\vert \times \vert \nabla_{\bm k} u_{\bm k}^s\rangle]
\end{equation}

\begin{eqnarray}
{\bm m}_{\bm k}^s=-\frac{e}{2\hbar}Im[\langle \nabla_{\bm k} u_k^s\vert \times (\mathcal{H}^s_0 - \epsilon_{\bm k}^s)\vert \nabla_{\bm k} u_k^s\rangle] 
\end{eqnarray}
where $\vert u_k^s\rangle$ satisfies the equation $ \mathcal{H}^s_0(\bm k)\vert u_k^s\rangle= \epsilon_{\bm k}^s\vert u_k^s\rangle $.\\

The general expressions for BC and OMM for WSMs are \cite{nandy2021chiral}
\begin{eqnarray}
\bm{\Omega}_{\bm{k}}^s =-s\frac{\bm k}{2k^3}\label{BC}\\
\bm{m}_{\bm{k}}^s=-s\frac{e \mathit{v}_F\bm k}{2k^2}\label{OMM}
\end{eqnarray}

From the Eqs.(\ref{BC}) and (\ref{OMM}), the OMM is related to the BC \cite{xiao2010berry}.
\begin{equation}
\mathbf m_{\mathbf k}^{s}=\frac{e}{\hbar}\,
\epsilon_{\mathbf k}^{s}\boldsymbol{\Omega}_{\mathbf k}^{s}
\end{equation}

\noindent Since the OMM is closely related to the BC, the two quantities jointly determine the magneto-optical response of Weyl semimetals. Their competition forms the central physical mechanism investigated in the present work.

The two equations (\ref{EOMa}) and (\ref{EOMb}) can be decoupled to get

\begin{eqnarray}
\dot{\bm r}=\frac{1}{D}[\bm{\tilde{v}}_{ \bm{k}}^s+\frac{e}{\hbar}\bm{E}\times\bm{\Omega}_{\bm{k}}^s
+\frac{e}{\hbar}(\bm{\tilde{v}}_{ \bm{k}}^s\cdot \bm{\Omega}_{\bm{k}}^s )\bm{B} ]\label{eomdecouplea}\\
\dot{\bm k}=\frac{1}{\hbar D}[-e\bm{E}-e\bm{\tilde{v}}_{ \bm{k}}^s\times \bm{B}-\frac{e^2}{\hbar}(\bm{E}.\bm{B}) \bm{\Omega}_{\bm{k}}^s]\label{eomdecoupleb}
\end{eqnarray}
 
\noindent where the factor D=1+$\frac{e}{\hbar}(\bm{\Omega}_{\bm{k}}^s\cdot\bm{B})$ modifies the phase space volume  \cite{xiao2005berry}. These equations contain three distinct contributions to carrier dynamics: $\bm{\tilde{v}}_{ \bm{k}}^s$ which governs ordinary charge transport in the absence of Berry curvature. Anomalous velocity, proportional to $\bm{E}\times\bm{\Omega}_{\bm{k}}^s$ which produces intrinsic Hall-type responses even without Lorentz-force deflection. Magnetic-field correction, arising from $(\bm{\tilde{v}}_{ \bm{k}}^s\cdot \bm{\Omega}_{\bm{k}}^s )\bm{B}$ which reflects the chiral nature of Weyl fermions and plays an essential role in magnetoconductivity.\\

To investigate the optical response, we employ the semiclassical Boltzmann transport
equation within the relaxation-time approximation,\\
\begin{eqnarray}
\frac{\partial \tilde{f}^s}{\partial t}+\dot{\bm k}\cdot\frac{\partial \tilde{f}^s}{\partial \bm k}=-\frac{ \tilde{f}^s-\tilde{f}^s_0}{\tau}\label{SBE}
\end{eqnarray}
Here, $\tilde{f}^s(\tilde{\epsilon_{\bm k}^s})$ is the electron distribution function and $\tau$ is the relaxation time originating from the scattering of electrons by phonons, impurities, electrons and other lattice imperfections of the solids\cite{malic2011microscopic}. \\

The $\tilde{f}_{0}^{s}( \epsilon_{\bm k}^s)$ can be Taylor expanded at low magnetic field as
\cite{pellegrino2015helicons}

\begin{eqnarray}
\tilde{f}_{0}^{s}(\tilde{\epsilon_{\bm k}^s})=\tilde{f}_{0}^{s}( \epsilon_{\bm k}^s -\bm{m}_{\bm{k}}^s\cdot \bm{B})\nonumber\\
\simeq \tilde{f}_{0}^{s}( \epsilon_{\bm k}^s)-\bm{m}_{\bm{k}}^s\cdot \bm{B}\frac{\partial \tilde{f}_{0}^{s}( \epsilon_{\bm k}^s)}{\partial \epsilon_{\bm k}^s}
\end{eqnarray}

\noindent where $\tilde{f}_{0}^{s}( \epsilon_{\bm k}^s)=1/[e^{ (\epsilon_{\bm k}^s-\mu)/k_B T} +1]$ with $k_B$ the Boltzmann constant, T the temperature and $\mu$ the chemical
potential.\\

\noindent Eq.(\ref{SBE}) can be solved by expanding the distribution function perturbatively as a power series in the applied electric field,

\begin{equation}
\tilde{f}^{s}=\tilde{f}_{0}^{s}+\tilde{f}_{1}^{s} e^{-i\omega t}+\tilde{f}_{2}^{s}e^{-2i\omega t}+....\label{fexp}
\end{equation}

\noindent where $\tilde{f}_{1}^{s}$ and $\tilde{f}_{2}^{s}$ are the first- and second-order terms for $\bm{E}$ respectively. The electric current density can be calculated by

\begin{eqnarray}
\bm{j}=-\frac{e}{(2\pi)^3}\int d^3k D\dot{\bm r}\tilde{f}^{s} \label{curr_den_main}
\end{eqnarray}

Equation(\ref{curr_den_main}) computes the conductivity components under the combined influence of external electric and magnetic fields. The present analysis is based on the semiclassical Boltzmann approximation, which is applicable when the magnetic field is sufficiently weak that Landau quantization can be neglected.\\

\section{Linear response of tilted WSMs}\label{linear_response}
For linear electric field response, we retain only the first two terms of Eq.(\ref{fexp}) and substitute Eq.(\ref{eomdecouplea}) into Eq.(\ref{SBE})
\begin{eqnarray}
\frac{1}{\hbar D}[-e \textbf{E}-\frac{e^2}{\hbar}(\textbf{E}\cdot\textbf{B})\bm{\Omega}_{\bm{k}}^s]\cdot \frac{\partial \tilde{f}_0^s}{\partial \bm k} -i \omega \tilde{f}_1^s=-\frac{\tilde{f}_1^s}{\tau}
\end{eqnarray}
Solved for $\tilde{f}_{1}^{s}$, we obtain
\begin{equation}
\tilde{f}_1^s=\frac{\tau}{(1-i\omega \tau)}\frac{1}{\hbar D}[e \textbf{E}+\frac{e^2}{\hbar}(\textbf{E}\cdot\textbf{B})\bm \Omega_{\bm k}^s]\cdot \frac{\partial \tilde{f}_0^s}{\partial \bm k}\label{linearE}
\end{equation}

We expand Eq.(\ref{linearE}) up to the second order in the magnetic field and obtain
\begin{widetext}
\begin{eqnarray}\label{disf}
\tilde{f}_1^s&=&\frac{\tau}{(1-i\omega \tau)}\biggl[e \textbf{E}\cdot \bm{\mathit{v}}_{\bm k}^s\cdot \frac{\partial \tilde{f}_0^s}{\partial \epsilon_{\bm k}^s}-\frac{e^2}{\hbar}(\textbf{B}\cdot \bm\Omega_{\bm k}^s)(\textbf{E}\cdot \bm{\mathit{v}}_{\bm k}^s )\frac{\partial \tilde{f}_0^s}{\partial \epsilon_{\bm k}^s}+\frac{e^2}{\hbar}(\textbf{E}\cdot \textbf{B})(\bm\Omega_{\bm k}^s \cdot \bm{\mathit{v}}_{\bm k}^s )\frac{\partial \tilde{f}_0^s}{\partial \epsilon_{\bm k}^s}-\frac{e}{\hbar}\textbf{E}\cdot\frac{\partial}{\partial \bm k}\biggl(\textbf{m}_{\bm k}^s\cdot \textbf{B}\frac{\partial \tilde{f}_0^s}{\partial \epsilon_{\bm k}^s}\biggr)\nonumber\\
&-&\frac{e^3}{\hbar^2}(\textbf{B}\cdot \bm\Omega_{\bm k}^s)(\textbf{E}\cdot \textbf{B})(\bm\Omega_{\bm k}^s \cdot \bm{\mathit{v}}_{\bm k}^s )\frac{\partial \tilde{f}_0^s}{\partial \epsilon_{\bm k}^s}+\frac{e^3}{\hbar^2}(\textbf{B}\cdot \bm\Omega_{\bm k}^s)^2(\textbf{E}\cdot \bm{\mathit{v}}_{\bm k}^s )\frac{\partial \tilde{f}_0^s}{\partial \epsilon_{\bm k}^s}+\frac{e^2}{\hbar^2}(\textbf{B}\cdot \bm\Omega_{\bm k}^s)\textbf{E}\cdot \frac{\partial}{\partial \bm k}\biggl(\textbf{m}_{\bm k}^s\cdot \textbf{B}\frac{\partial \tilde{f}_0^s}{\partial \epsilon_{\bm k}^s}\biggr)\nonumber\\
&-&\frac{e^2}{\hbar^2}\bm\Omega_{\bm k}^s\cdot \frac{\partial}{\partial \bm k}\biggl(\textbf{m}_{\bm k}^s\cdot \textbf{B}\frac{\partial \tilde{f}_0^s}{\partial \epsilon_{\bm k}^s}\biggr)\biggr]
\end{eqnarray}
\end{widetext}

From Eqs.(\ref{eomdecouplea}) and Eq.(\ref{disf}), the expression for the current density at time t is given by
\begin{eqnarray}
\bm{j}_1&=&-\frac{e}{(2\pi)^3}\int d^3k \Bigl[\bm{\mathit{\tilde{v}}}_{\bm k}^s+\frac{e}{\hbar}(\Omega_{\bm k}^s \cdot \bm{\mathit{\tilde{v}}}_{\bm k}^s )\textbf{B}\Bigr]\tilde{f}_1^{s}\label{cur_den} \nonumber\\
&-&\frac{e^2}{(2\pi)^3\hbar}\int d^3k \textbf{E}\times \bm\Omega_{\bm k}^s \tilde{f}_0^s\label{cond_eqn}
\end{eqnarray}

The above equation can be expressed in the frequency space $\omega$ as
\begin{equation}
j_a(\omega)=\sigma_{ab}(\omega)E_b(\omega)
\end{equation}

\noindent where $\sigma_{ab}(\omega)$ is the frequency $\omega$ dependent conductivity. It is known that the single contribution from the group velocity $\bm{\mathit{\tilde{v}}}_{\bm k}^s$ or the BC $\bm{\Omega}_{\bm{k}}^s$ forms the conventional longitudinal or Hall conductivities. In the presence of the magnetic field, the conductivity $\sigma(\omega)$ consists of the coupling terms between the group velocity  $\bm{\mathit{\tilde{v}}}_{\bm k}^s$ and the BC $\bm{\Omega}_{\bm{k}}^s$ besides the conventional ingredients [see Eq.(\ref{cond_eqn})]and play a crucial role in the electron transport.

\subsection{Calculations of longitudinal conductivity components without magnetic field}
Substituting Eq.(\ref{disf}) without $ \textbf{B} $ terms into the first term of Eq.(\ref{cur_den}), we get the longitudinal conductivities components
\begin{equation}
\sigma_{ab}^{(0)}(\omega)=\frac{\tau}{(1-i\omega \tau)}\frac{e^2}{(2\pi)^3}\int d^3k \mathit{v}_a^s \mathit{v}_b^s\Bigl(-\frac{\partial f_0^s}{\partial \epsilon_{\bm k}^s}\Bigr)
\end{equation}

\onecolumngrid\
\begin{center}\
\begin{figure}
\begin{tabular}{cc}
\includegraphics[width=0.5\linewidth]{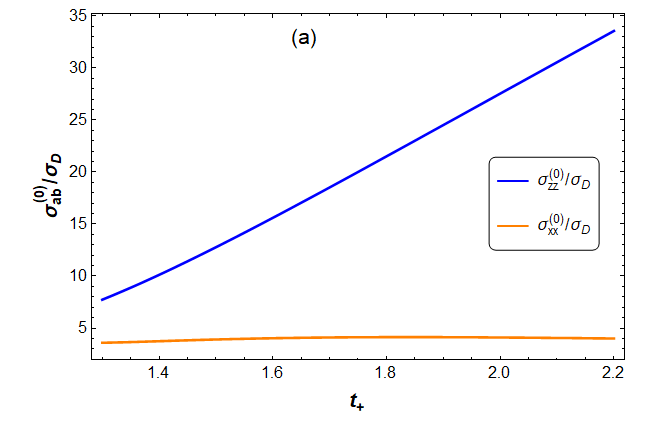} \
\includegraphics[width=0.5\linewidth]{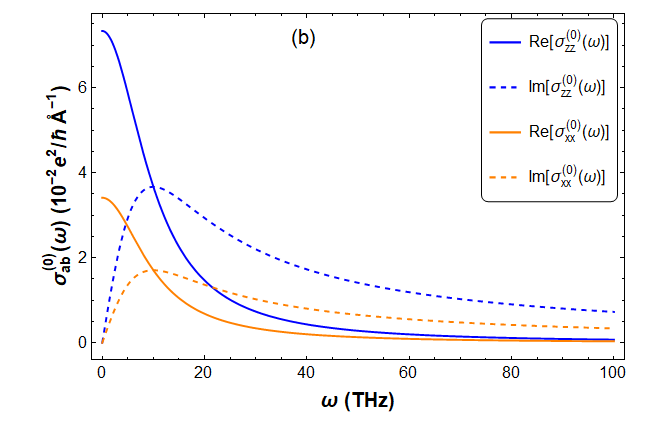}  
\end{tabular}
\caption{(a) The tilt $t_+$ dependence of the Drude conductivity(in units of $\sigma_D$). Unlike the case of type-I WSM, $\sigma_{zz}^0$ increases more rapidly than $\sigma_{xx}^0$.(b) The frequency dependence of the Drude conductivity at $t_s=1.3$. The other parameters are taken as $\tilde{\Lambda}_k=4$, $v_F=4.13\times 10^5$ m/s, $\mu$=0.1 eV and $\tau=10^{-13}s$.}
\label{fig_cond_noB_tilt}\
\end{figure}\
\end{center}\
\twocolumngrid\

At T=0 K, $-\frac{\partial f_0^s}{\partial \varepsilon_{\bm k}^s}=\delta(\varepsilon_{\bm k}^s-\mu)$, we get\\

\begin{eqnarray}
\sigma_{zz}^0(\omega)&=&\frac{3\sigma_D}{2\lvert t_s \rvert^5} {\bigg[3-t_s^2+(t_s^2-1)^2\tilde{\Lambda}_k^2+\ln[(t_s^2-1)\tilde{\Lambda}_k^2}]\bigg]\nonumber\\ 
\end{eqnarray}
\begin{eqnarray}
\sigma_{xx}^0(\omega)&=&\frac{3\sigma_D}{4\lvert t_s \rvert^5}\bigg[\frac{3-t_s^2}{t_s^2-1}+(t_s^2-1)\tilde{\Lambda}_k^2-\ln[(t_s^2-1)\tilde{\Lambda}_k^2]\bigg]\nonumber\\
\end{eqnarray}
\label{cond_noB_single_xx}
\noindent where
\begin{equation}
\sigma_D=\frac{e^2\tau\mu^2}{(1-i\omega \tau)6\pi^2 \hbar^3\mathit{v_F}}
\end{equation}
is the Drude frequency complex conductivity and $\Lambda_k/k_F\equiv \tilde{\Lambda}_k $ is the ultraviolet momentum cutoff along the radial direction in spherical polar co-ordinates to avoid infinity in conductivity calculations(see Appendix A). This cutoff dependence reflects the continuum description adopted for the overtilted Weyl cone and should not be interpreted as a universal property of Weyl semimetals. In realistic materials, the finite Brillouin zone naturally regularizes these terms while preserving the qualitative dependence on the tilt parameter. In a tight-binding crystal with lattice spacing a, one expects $\Lambda_k \sim \mathcal{O}(\pi/a)$. Unlike the case of type-I WSM, for a type-II WSM we find that $\sigma_{zz}^{(0)}$ is larger and more sensitive to $t_s$ as compared to $\sigma_{xx}^{(0)}$[see Fig.(\ref{fig_cond_noB_tilt}a)] \cite{das2019linear}. The frequency dependence of conductivity $\sigma_{aa}^{(0)}$(a=x,z) exhibits a Drude-type behavior, as shown in Fig.(\ref{fig_cond_noB_tilt}b).\\


\subsection{Calculations of conductivity components linear in magnetic field}
Substituting Eq.(\ref{disf}) with $ \textbf{B} $ terms up to first order into the first term of Eq.(\ref{cur_den}), we get conductivities components
\begin{equation} 
\sigma_{ab}^{(B)}(\omega)=\sigma_{ab}^{(B,\Omega)}(\omega)+\sigma_{ab}^{(B,m)}(\omega)
\label{cond_linearB}
\end{equation}
where
\begin{widetext}
\begin{eqnarray}
\sigma_{ab}^{(B,\Omega)}(\omega)&=&\frac{\tau}{\hbar(1-i\omega \tau)}\frac{e^3}{(2\pi)^3}\int d^3k [(\mathit{v}_a^s B_b+\mathit{v}_b^s B_a)(\bm\Omega_{\bm k}^s \cdot \bm{\mathit{v}}_{\bm k}^s )-\mathit{v}_a^s \mathit{v}_b^s(\bm\Omega_{\bm k}^s \cdot \textbf{B})]\Bigl(-\frac{\partial f_0^s}{\partial \epsilon_{\bm k}^s}\Bigr)\label{cond_B}\\
\sigma_{ab}^{(B,m)}(\omega)&=&\frac{\tau}{\hbar(1-i\omega \tau)}\frac{e^2}{(2\pi)^3}\int d^3k\Bigl[\frac{\partial \mathit{v}_a^s }{\partial k_b}(\textbf{m}_{\bm k}^s\cdot \textbf{B})-\frac{\partial (\textbf{m}_{\bm k}^s\cdot \textbf{B})}{\partial k_a}\mathit{v}_b^s \Bigr]\Bigl(-\frac{\partial f_0^s}{\partial \epsilon_{\bm k}^s}\Bigr)\label{condcomp}
\end{eqnarray}
\end{widetext}

We can easily check from Eqs.(\ref{cond_B}) and (\ref{condcomp}) that $\sigma_{ab}^{(B)}(\omega)=\sigma_{ba}^{(B)}(\omega)$. For the conductivity contributions considered here, we find that they vanish in the absence of a tilt term (\(t_s=0\)) \cite{gao2022suppression,morimoto2016semiclassical}. In contrast, a finite tilt (\(t_s\neq 0\)) gives rise to nonzero contributions. The first term in Eq.(\ref{cond_linearB}) is related to the BC and is reported previously
\cite{das2019linear}. The second term in Eq.(\ref{cond_linearB}) is our newly obtained result and stems from the contribution of the OMM.
In the following, the detailed analysis of Eqs.(\ref{cond_B})  and (\ref{condcomp}) is given by considering the magnetic field $\textbf{B}$ perpendicular and parallel to the tilt direction $\hat{\bm{t}}_s$.\\

\onecolumngrid\
\begin{center}\
\begin{figure}
\begin{tabular}{ccc}
\includegraphics[width=0.5\linewidth]{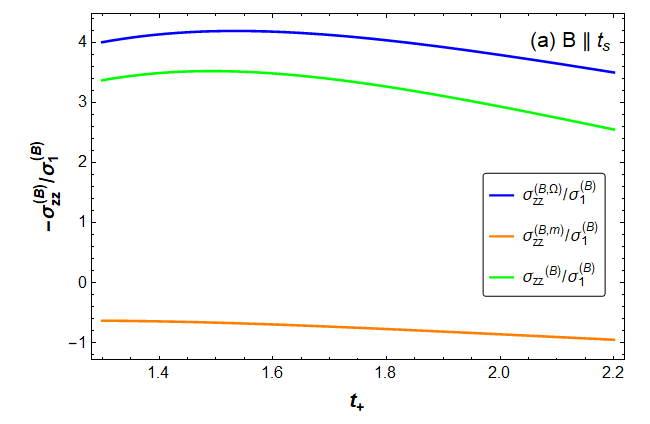} \
\includegraphics[width=0.5\linewidth]{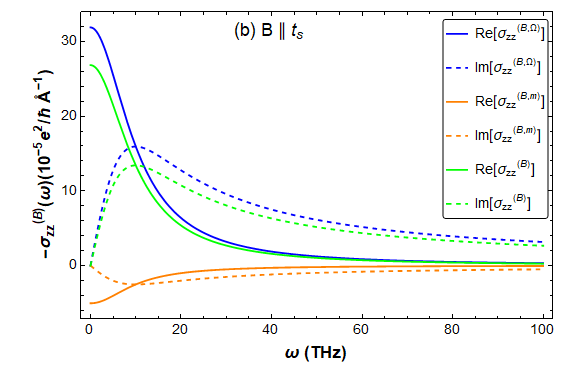} \    
\end{tabular}
\caption{(a) The tilt($t_+$) dependence of the B-linear conductivity $\sigma_{zz}^{(B)}$ for the case of $\textbf{B}\Vert \hat{\bm t}_s$ at B=1 T. (b) The frequency dependence of the B-linear magnetoconductivity  at $t_s=1.3$. All the other parameters are identical to those in Fig.(\ref{fig_cond_noB_tilt}).}
\label{cond_linearB_pll_zz_tilt}\
\end{figure}\
\end{center}\
\twocolumngrid\

\begin{figure}
\includegraphics[scale=0.38]{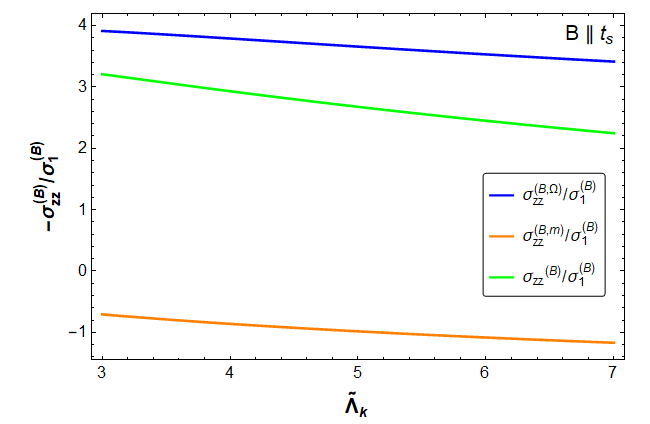}
\caption{ The radial momentum cutoff $\tilde{\Lambda}_k$ dependence of the B-linear conductivity $\sigma_{zz}^{(B)}$ for the case of $\textbf{B}\Vert \hat{\bm t}_s$ at B=1 T and $t_+=2.0$. All the other parameters are identical to those in Fig.(\ref{fig_cond_noB_tilt}).}
\label{cond_linearB_zz_lambda}\
\end{figure}\

Case-I When $\textbf{B}\Vert \hat{\bm t}_s \Vert \hat{\bm z} $\\
Owing to the rotational symmetry around the tilt axis, only the longitudinal conductivity
components remain finite while all transverse components vanish.
 
\begin{widetext}
\begin{eqnarray}\label{Berry_linearB}
\sigma_{zz}^{(B,\Omega)}&=&\frac{\sigma_1^{(B)}s}{3t_s^4}\sgn(t_s) \bigg[-4\tilde{\Lambda}_k^{-3}+3(-3+2t_s^2)\tilde{\Lambda}_k^{-2}+36(-1+t_s^2)\Lambda_k^{-1}+\big(-11+\lvert t_s \rvert \big\{12+t_s(21-4\lvert t_s \rvert(5+3\lvert t_s \rvert))\big\}\big)\nonumber\\&+&3(t_s^2-1)^2\Big\{3\ln(-1+\lvert t_s \rvert)-\ln(1+\lvert t_s \rvert)+2\ln\tilde{\Lambda}_k\Big\}\bigg]
\end{eqnarray}

\begin{eqnarray}\label{cond_xx_B_typeII}
\sigma_{xx}^{(B,\Omega)}&=&\frac{\sigma_1^{(B)}s}{6t_s^4}\sgn(t_s) \bigg[-9\tilde{\Lambda}_k^{-2}+(-11+9t_s^2)-3(-1+t_s^2)\ln[(t_s^2-1)\tilde{\Lambda}_k^2]\bigg]
\end{eqnarray}
\end{widetext}

Via the similar calculation, Eq.(\ref{condcomp}) becomes

\begin{eqnarray}\label{orbital_B}
\sigma_{zz}^{(B,m)}(\omega)&=&\frac{ \sigma_1^{(B)}s}{6t_s^4}\sgn(t_s) \bigg[\tilde{\Lambda}_k^{-2}(-9+6t_s^2)+(-11+21t_s^2\nonumber\\&-&6t_s^4)+3(-1+t_s^2)^2\ln[(t_s^2-1)\ln\tilde{\Lambda}_k^2]\bigg]
\end{eqnarray}
\begin{eqnarray}
\sigma_{xx}^{(B,m)}(\omega)&=&\frac{\sigma_1^{(B)}s}{6t_s^4}\sgn(t_s) \bigg[-2\tilde{\Lambda}_k^{-3}+18(-1+t_s^2)\tilde{\Lambda}_k^{-1}\nonumber\\&+&2\lvert t_s \rvert(3-8t_s^2)-3(-1+3t_s^2)\ln\frac{\lvert t_s \rvert-1}{\lvert t_s \rvert+1}\bigg]\nonumber\\
\label{orbital_B_typeII}
\end{eqnarray}

\noindent where 
\begin{equation}
\sigma_1^{(B)}=\frac{e^3\tau B\mathit{v_F}}{(1-i\omega \tau)8\pi^2\hbar^2}
\end{equation}

The total linear-B contribution of two nodes having parallel tilt becomes zero. But for nodes having opposite tilt, as is usually the case, the total contribution is nonzero. Fig.(\ref{cond_linearB_pll_zz_tilt}a) demonstrates that the OMM suppresses the $\sigma_{zz}^{(B)}$ conductivity over the entire range of tilt parameters. This suppression originates from the negative contribution of the OMM to the effective carrier velocity. Thus, the total linear B conductivity $\sigma_{zz}^{(B)}$  nonmonotonically decreases with tilt in the type-II regime rather than decreasing in type-I regime \cite{gao2022suppression}. Figure (\ref{cond_linearB_pll_zz_tilt}b) illustrates the frequency dependence of the magnetoconductivity. We observe that $\sigma^{(B)}(\omega)$ is approximately two orders of magnitude smaller than $\sigma^{(0)}(\omega)$ over the frequency range considered.\\
 
From Eqs.(\ref{Berry_linearB})-(\ref{orbital_B_typeII}) unlike the case of type-I WSM, there is Fermi energy dependence that is tied to the momentum cutoff $\tilde{\Lambda}_k$ in the above linear B conductivity calculations. To examine whether the conclusions regarding the BC and the OMM are sensitive to our choice of $\tilde{\Lambda}_k=4 $, we have repeated the calculations for $\tilde{\Lambda}_k $ from 3 to 7 for $\sigma_{zz}^{(B)}(\omega)$ as shown in Fig.(\ref{cond_linearB_zz_lambda}). Although the absolute values of the conductivity components change with the cutoff, the qualitative OMM-induced enhancement or suppression remains unchanged. In particular, the BC and OMM corrections retain the same trend over the cutoff range considered, demonstrating that the main conclusions concerning the role of the BC and OMM, including its enhanced importance in the type-II regime, are not artifacts of the particular choice $\tilde{\Lambda}_k$. \\

The behavior of the conductivity component $\sigma^{(B)}_{xx}$ differs from that of $\sigma^{(B)}_{zz}$. For type-I Weyl semimetals, the OMM reduces the longitudinal conductivity
\cite{gao2022suppression}. In contrast, for type-II Weyl semimetals, the OMM contributes positively, leading to a systematic enhancement of the total conductivity. This contrasting behavior clearly demonstrates that the role of the OMM depends not only on its intrinsic coupling to the magnetic field but also on the topology of the Fermi surface. The Lifshitz transition therefore fundamentally changes the influence of the OMM on magneto-optical transport. Fig.(\ref{cond_linearB_pll_xx_tilt}a) and Fig.(\ref{cond_linearB_pll_xx_tilt}b) show the tilt and frequency dependence of $\sigma_{xx}^{(B)}$.

\onecolumngrid\
\begin{center}\
\begin{figure}
\begin{tabular}{ccc}
\includegraphics[width=0.5\linewidth]{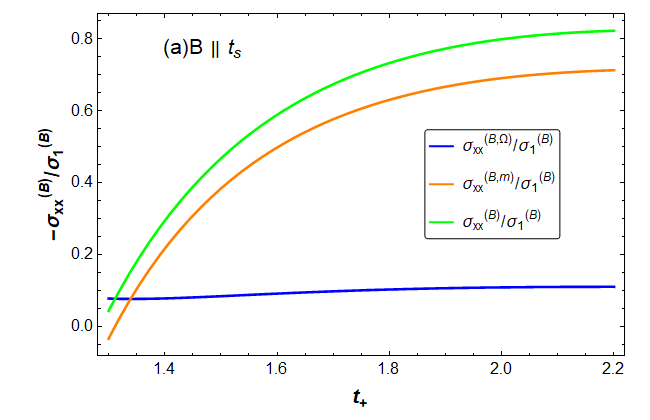} \
\includegraphics[width=0.5\linewidth]{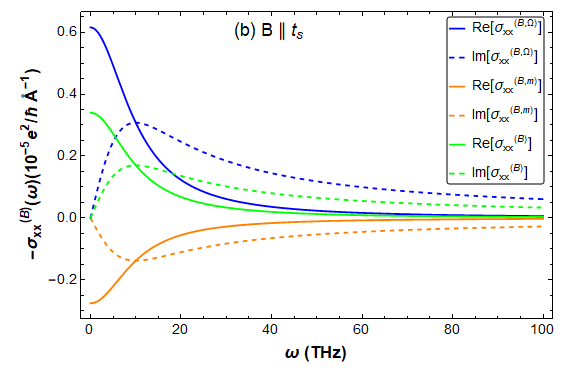} \    
\end{tabular}
\caption{(a) The tilt $t_+$ dependence of the B-linear conductivity $\sigma_{xx}^{(B)}$ for the case of $\textbf{B}\Vert \hat{\bm t}_s$ at B=1 T.(b) The frequency dependence of the magnetoconductivity at $t_s=1.3$. All the other parameters are identical to those in Fig.(\ref{fig_cond_noB_tilt}).}
\label{cond_linearB_pll_xx_tilt}\
\end{figure}\
\end{center}\
\twocolumngrid\

Case-II For $\textbf{B}\perp \hat{\bm t}_s (\Vert \hat{\bm z}) $\\ 

 We can represent the magnetic field in the x-y plane as $\textbf{B}=B(\cos\gamma \hat{\bf i} + \sin \gamma \hat{\bf j})$, where $\gamma$ is the angle between the magnetic field $ \textbf{B} $ and the x-axis. Unlike the parallel-field configuration, the perpendicular geometry allows the BC to generate a finite planar Hall current even in the absence of Lorentz-force deflection. The OMM subsequently modifies this Hall response through the magnetic-field-induced correction to the quasiparticle energy. In this configuration, the longitudinal conductivity components vanish, yielding the following expressions for the planar Hall conductivities \cite{gao2022suppression, das2019linear}
\begin{equation}
\sigma_{xz}^{(B)}(\omega)=[\sigma^{(B,\Omega)}(\omega)+\sigma^{(B,m)}(\omega)]\cos \gamma
\end{equation}

\begin{equation}
\sigma_{yz}^{(B)}(\omega)=[\sigma^{(B,\Omega)}(\omega)+\sigma^{(B,m)}(\omega)]\sin \gamma
\end{equation}

\begin{widetext}
\begin{eqnarray}
\sigma^{(B,\Omega)}_{xz}(\omega)=\frac{\tau}{\hbar
(1-i\omega \tau)}\frac{e^3}{(2\pi)^3}\int d^3k [\mathit{v}_z^s B \cos \gamma (\mathit{v}_y^s\Omega_{ky}+\mathit{v}_z^s\Omega_{kz})-\mathit{v}_z^s B \sin \gamma \mathit{v}_x^s\Omega_{ky}]\Bigl(-\frac{\partial f_0^s}{\partial \epsilon_{\bm k}^s}\Bigr)
\end{eqnarray}

\begin{eqnarray}
\sigma^{(B,m)}_{xz}(\omega)=\frac{\tau}{\hbar(1-i\omega \tau)}\frac{e^2}{(2\pi)^3}\int d^3k \Bigl[ B \cos \gamma \Bigl(\frac{\partial \mathit{v}_x^s}{\partial k_z}m_{kx}^s-v_z^s\frac{\partial \mathit{m}_{kx}^s}{\partial k_z}\Bigr)- B \sin \gamma \Bigl(\frac{\partial \mathit{v}_x^s}{\partial k_z}m_{ky}^s-v_z^s\frac{\partial \mathit{m}_{ky}^s}{\partial k_x}\Bigr)\Bigr]\Bigl(-\frac{\partial f_0^s}{\partial \epsilon_{\bm k}^s}\Bigr)
\end{eqnarray}

The second term of both the above equations contributes to zero.

\begin{eqnarray}
\sigma^{(B,\Omega)}&=&\frac{\sigma_1^{(B)}s}{6t_s^4}\sgn(t_s) \biggl[-2\tilde{\Lambda}_k^{-3}+6(-3+2t_s^2)\tilde{\Lambda}_k^{-2}+18(-1+t_s^2)\tilde{\Lambda}_k^{-1}-2(1+\lvert t_s \rvert)\big\{11+\lvert t_s \rvert(-14+\lvert t_s \rvert(-7+12\lvert t_s \rvert))\big\}\nonumber\\&+&3(-1+t_s^2)^2\Big\{3\ln(-1+\lvert t_s\rvert)+\ln(1+\lvert t_s\rvert)-4\ln\tilde{\Lambda}_k\Big\}\bigg]
\end{eqnarray}
\begin{eqnarray}\label{linear_B_cond_perp}
\sigma^{(B,m)}&=&\frac{\sigma_1^{(B)}s}{6t_s^4}\sgn(t_s) \bigg[2\Big\{\tilde{\Lambda}_k^{-3}-3(-3+t_s^2)\tilde{\Lambda}_k^{-2}-3(-3+t_s^2)\tilde{\Lambda}_k^{-1}+\big\{11+\lvert t_s \rvert(-3+t_s(-12+\lvert t_s \rvert(2+3\lvert t_s \rvert)))\big\}\Big\}\nonumber\\&+&3(-1+t_s^2)\Big\{3\ln(-1+\lvert t_s \rvert)+\ln(1+\lvert t_s \rvert)-4\ln\tilde{\Lambda}_k\Big\}\bigg]
\end{eqnarray}
\end{widetext}

The linear dependence of the rest of the conductivities is zero, i.e., $\sigma_{xy}$=$\sigma_{xx}$=$\sigma_{yy}$= $\sigma_{zz}=0$. From Eqs.(\ref{Berry_linearB}) to (\ref{linear_B_cond_perp}), we notice that the B-linear magnetoconductivity is dependent of Fermi energy $\mu$ that tied to the momentum cutoff $\tilde{\Lambda}_k$ and the odd function of $t_s$ \cite{gao2022suppression}. According to the Nielsen-Ninomiya theorem \cite{nielsen1981absence,nielsen1981absence}, the Weyl nodes with opposite chirality always appear in pairs. The total magnetoconductivity of the system is the sum of all the Weyl nodes. Therefore, in  the equal-tilt configuration $t_{+} =t_{-}$, the contribution of the Weyl node to the magnetoconductivity has the opposite sign for the opposite chirality, giving rise to $\sigma_{ab}(\omega)$ = 0. Whereas, for opposite-tilt configuration $t_{+} =-t_{-}$, each Weyl node produces an identical contribution to the magnetoconductivity and so the nonzero magnetoconductivity emerges for this case. Fig.(\ref{cond_linearB_tilt}a) and (\ref{cond_linearB_tilt}b) show the tilt and frequency variations of linear B planar Hall conductivity components. We find that $\sigma^{(B)}(\omega)$ is about two order of magnitude smaller than $\sigma^{(0)}(\omega)$. The suppression takes place with the tilt parameter in the opposite manner in both the type-I and type-II phases\cite{gao2022suppression}. Experimentally, candidate type-II Weyl semimetals such as WTe$_2$ and MoTe$_2$ exhibit pronounced BC-induced linear planar Hall effects \cite{chen2018planar, li2019anisotropic}. However, to the best of our knowledge, no experiment has yet systematically isolated the OMM contribution from the BC contribution to the linear magnetoconductivity of tilted Weyl semimetals. Our results therefore provide experimentally testable predictions for identifying OMM-induced modifications to the magnetoconductivity, rather than claiming that such effects have already been uniquely resolved in existing measurements.\\

\onecolumngrid\
\begin{center}\
\begin{figure}
\begin{tabular}{ccc}
\includegraphics[width=0.5\linewidth]{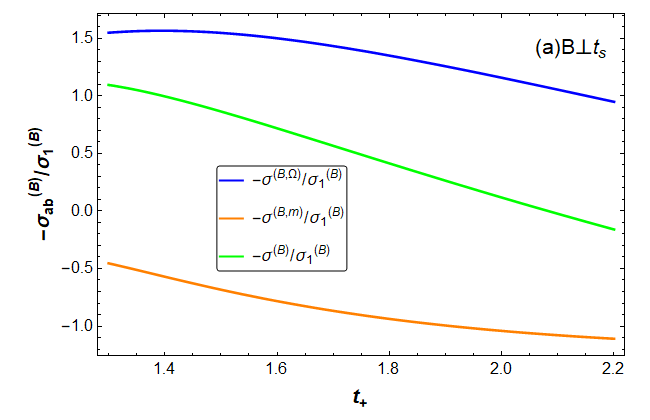} \
\includegraphics[width=0.5\linewidth]{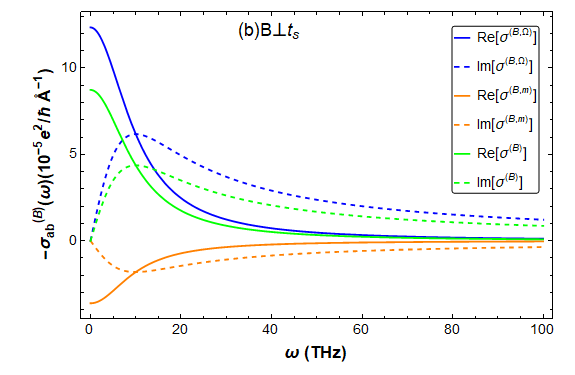} \    
\end{tabular}
\caption{(a) The tilt $t_+$ dependence of the linear-B planar Hall conductivity(in unitless) for the case of $\textbf{B}\perp \hat{\bm t}_s $ at B=1 T.(b) The frequency dependence of conductivity at $t_s=1.3$. All the other parameters are identical to those in Fig.(\ref{fig_cond_noB_tilt}).}
\label{cond_linearB_tilt}\
\end{figure}\
\end{center}\
\twocolumngrid\

\subsection{Calculations of quadratic-B contribution to the conductivity $\sigma_{ab}^{(B^2)}$}

Substituting Eq.(\ref{disf}) with $ \textbf{B} $ terms up to second order into the first term of Eq.(\ref{cur_den}), we get conductivities components

\begin{equation}
\sigma_{ab}^{(B^2)}(\omega)=\sigma_{ab}^{(B^2,\Omega)}(\omega)+\sigma_{ab}^{(B^2,m)}(\omega)
\end{equation}

where

\begin{widetext}
\begin{equation}
\begin{split}
\sigma_{ab}^{(B^2,\Omega)}(\omega)=& \frac{\tau}{\hbar^2(1-i\omega \tau)}\frac{e^4}{(2\pi)^3}\int d^3k [\mathit{v}_a^s \mathit{v}_b^s(\bm\Omega_{\bm k}^s \cdot \textbf{B})^2-(\mathit{v}_a^s B_b+\mathit{v}_b^s B_a)(\bm\Omega_{\bm k}^s \cdot \bm{\mathit{v}}_{\bm k}^s )(\bm\Omega_{\bm k}^s \cdot \textbf{B})\\& + B_a B_b (\bm\Omega_{\bm k}^s \cdot \bm{\mathit{v}}_{\bm k}^s )^2]\Bigl(-\frac{\partial f_0^s}{\partial \epsilon_{\bm k}^s}\Bigr)
\end{split}
\end{equation}

\begin{equation}
\begin{split}
\sigma_{ab}^{(B^2,m)}(\omega)=&\frac{\tau}{\hbar^2(1-i\omega \tau)}\frac{e^3}{(2\pi)^3}\int d^3k\Bigl[\frac{\partial}{\partial \textbf{k}}\cdot[\mathit{v}_a^s B_b \bm\Omega_{\bm k}^s](\textbf{m}_{\bm k}^s\cdot \textbf{B})-\frac{\partial [\mathit{v}_a^s(\bm\Omega_{\bm k}^s \cdot \textbf{B}) ]}{\partial k_b}(\textbf{m}_{\bm k}^s\cdot \textbf{B})+\frac{\partial [B_a^s(\bm\Omega_{\bm k}^s \cdot \bm{\mathit{v}}_{\bm k}^s ) ]}{\partial k_b}(\textbf{m}_{\bm k}^s\cdot \textbf{B})+ \\& \frac{\partial (\textbf{m}_{\bm k}^s\cdot \textbf{B})}{\partial k_a}(\bm\Omega_{\bm k}^s \cdot \textbf{B})\mathit{v}_b^s-\frac{\partial (\textbf{m}_{\bm k}^s\cdot \textbf{B})}{\partial k_a}(\bm\Omega_{\bm k}^s \cdot \bm{\mathit{v}}_{\bm k}^s )B_b^s -\frac{\partial^2 (\textbf{m}_{\bm k}^s\cdot \textbf{B})}{e \partial k_a \partial k_b}(\textbf{m}_{\bm k}^s\cdot \textbf{B})-B_a\frac{\partial (\textbf{m}_{\bm k}^s\cdot \textbf{B})}{\partial \bm k}\cdot \bm\Omega_{\bm k}^s \mathit{v}_b^s \Bigr]\Bigl(-\frac{\partial f_0^s}{\partial \epsilon_{\bm k}^s}\Bigr)
\end{split}
\end{equation}
\end{widetext}

\noindent where $\sigma_{ab}^{(B^2,\Omega)}(\omega)$ and $\sigma_{ab}^{(B^2,m)}(\omega)$
 are BC $\bm{\Omega}_{\bm k}^s$ and the OMM $\textbf{m}_{\bm k}^s$ dependent conductivities respectively. Consider the following two cases.\\
\onecolumngrid\
\begin{center}\
\begin{figure}
\begin{tabular}{ccc}
\includegraphics[width=0.5\linewidth]{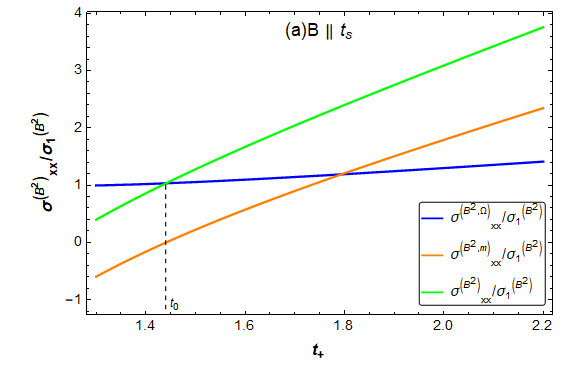} \
\includegraphics[width=0.5\linewidth]{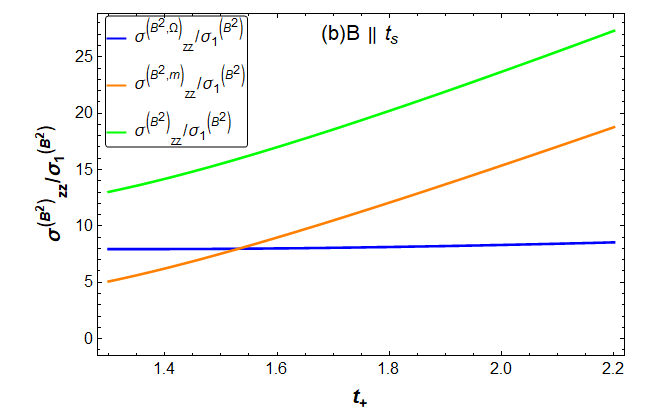} \    
\end{tabular}
\caption{(a) The tilt $t_+$ dependence of the quadratic B longitudinal conductivity $\sigma_{xx}^{(B^2)}$.(b) The tilt $t_+$ dependence of the quadratic B longitudinal conductivity $\sigma_{zz}^{(B^2)}$ for the case of  $\bm B \parallel \hat{\bm t}_s$ at B=1 T. All the other parameters are identical to those in Fig.(\ref{fig_cond_noB_tilt}).}
\label{quadraticB_cond_parallel}\
\end{figure}\
\end{center}\
\twocolumngrid\
Case-I When $\textbf{B}\Vert \hat{\bm t}_s \Vert \hat{\bm z} $\\
 In this case, Hall conductivity components are zero and one can get the following expressions for longitudinal components.

\begin{eqnarray}\label{cond_quadB_zz_typeI}
\sigma_{zz}^{(B^2,\Omega)}(\omega)&=&\frac{\sigma_1^{(B^2)}}{2\lvert t_s \rvert^5}\bigg[-5\tilde{\Lambda}_k^{-6}+15(-3+t_s^2)\tilde{\Lambda}_k^{-4}-15 (-1\nonumber\\ &+&t_s^2)^2\tilde{\Lambda}_k^{-2}+\Big\{1+5t_s^2(-1+3t_s^2+t_s^4)\Big\}\bigg]
\end{eqnarray}
\begin{eqnarray}
\sigma_{xx}^{(B^2,\Omega)}(\omega)&=&\frac{\sigma_1^{(B^2)}}{8\lvert t_s \rvert^5}\bigg[10\tilde{\Lambda}_k^{-6}-15(-6+t_s^2)\tilde{\Lambda}_k^{-4}-30(-1\nonumber\\&+&t_s^2)\tilde{\Lambda}_k^{-2}+(-2+5(t_s^2+t_s^6))\bigg]
\end{eqnarray}

Considering the effect of the OMM, we have


\begin{widetext}
\begin{eqnarray}
\sigma_{zz}^{(B^2,m)}(\omega)&=&\frac{\sigma_1^{(B^2)}}{16\lvert t_s \rvert^5}\bigg[320\tilde{\Lambda}_k^{-6}+384\tilde{\Lambda}_k^{-5}+60(48-7t_s)\tilde{\Lambda}_k^{-4}-160(-4+t_s^2)\tilde{\Lambda}_k^{-3}+15(64-56t_s^2-3t_s^4)\tilde{\Lambda}_k^{-2}\nonumber\\&+&\Big\{-64+140t_s^2+t_s^4\big\{35+\lvert t_s \rvert(-374+295\lvert t_s \rvert)\big\}\Big\}\bigg]
\end{eqnarray}
\begin{eqnarray}\label{cond_quadB_xx_typeII}
\sigma_{xx}^{(B^2,m)}(\omega)&=&\frac{-\sigma_1^{(B^2)}}{2\lvert t_s \rvert^5}\bigg[20\tilde{\Lambda}_k^{-6}+24\tilde{\Lambda}_k^{-5}-15(-12+t_s^2)\tilde{\Lambda}_k^{-4}+40\tilde{\Lambda}_k^{-3}-30(-2+t_s^2)\tilde{\Lambda}_k^{-2}+(-4+5t_s^2+6t_s^5-5t_s^6)\bigg]\nonumber\\
\end{eqnarray}
\end{widetext}

\noindent where
\begin{equation}
\sigma_1^{(B^2)}=\frac{e^4 \tau  B^2 v_F^3}{8\pi^2\hbar(1-i\omega \tau)15\mu^2}
\end{equation}

For type-II WSMs, the quadratic-in-magnetic-field conductivity contributions arising from the BC and OMM do not cancel each other. Consequently, the total conductivity \(\sigma_{xx}^{(B^2)}(\omega)\) is enhanced for tilt parameters \(t_0>1.44\) due to the OMM contribution, as shown in Fig.(\ref{quadraticB_cond_parallel}a). Furthermore, the conductivity component \(\sigma_{zz}^{(B^2)}(\omega)\) increases with the tilt parameter \(t_+\), as illustrated in Fig. (\ref{quadraticB_cond_parallel}b), owing to the positive contribution from the OMM. In contrast, for type-I WSMs, the quadratic-in-magnetic-field contributions to \(\sigma_{xx}^{(B^2,\Omega)}(\omega)\) and \(\sigma_{xx}^{(B^2,m)}(\omega)\), arising from the BC and OMM, respectively, are equal in magnitude but opposite in sign. These contributions therefore cancel, resulting in a vanishing total quadratic conductivity \(\sigma_{xx}^{(B^2)}(\omega)\)  \cite{gao2022suppression}. However, the behavior of \(\sigma_{zz}^{(B^2)}(\omega)\) differs: the total conductivity is enhanced for \(t_0>0.78\)\cite{gao2022suppression}.\\

Case-II For $\textbf{B}\perp \hat{\bm t}_s (\Vert \hat{\bm z}) $\\
In this case, both longitudinal components as well as planer Hall conductivities are non-zero. The longitudinal components have the following expressions

\begin{widetext}
\begin{eqnarray}
\sigma_{xx}^{(B^2,\Omega)}(\omega)&=&\frac{\sigma_1^{(B^2)}}{16\lvert t_s \rvert^5}\bigg[\Big\{-55\tilde{\Lambda}_k^{-6}+48(-4+t_s^2)\tilde{\Lambda}_k^{-5}-15(33-27t_s^2+4t_s^4)\tilde{\Lambda}_k^{-4}-80(4-5t_s^2+t_s^4)\tilde{\Lambda}_k^{-3}-165(-1+t_s^2)^2\tilde{\Lambda}_k^{-2}\nonumber\\&+&(1+t_s^2)\big\{11-50t_s^2+t_s^4(115+4\lvert t_s \rvert(8+15\lvert t_s \rvert))\big\}\Big\}\cos^2\gamma+\Big\{-5\tilde{\Lambda}_k^{-6}+15(-3+t_s^2)\tilde{\Lambda}_k^{-4}-15(-1+t_s^2)^2\tilde{\Lambda}_k^{-2}\nonumber\\&+&\big\{1+5t_s^2(-1+3t_s^2+t_s^4)\big\}\Big\}\sin^2\gamma\bigg]\\
\sigma_{zz}^{(B^2,\Omega)}(\omega)&=&\frac{\sigma_1^{(B^2)}}{8\lvert t_s \rvert^5}\bigg[10\tilde{\Lambda}_k^{-6}+15(6-7t_s^2+t_s^4)\tilde{\Lambda}_k^{-4}-30(-1+t_s^2)^3\tilde{\Lambda}_k^{-2}+\big\{-2+11t_s^2-25t_s^4+65t_s^6+15t_s^8\big\}\bigg]
\end{eqnarray}
\end{widetext}

The off-diagonal components of conductivities

\begin{widetext}
\begin{eqnarray}
\sigma_{xy}^{(B^2,\Omega)}(\omega)&=&\frac{\sigma_1^{(B^2)}}{8\lvert t_s \rvert^5}\bigg[-25\tilde{\Lambda}_k^{-6}+24(-4+t_s^2)\tilde{\Lambda}_k^{-5}-15(-5+t_s^2)(-3+2 t_s^2)\tilde{\Lambda}_k^{-4}-40(4-5t_s^2+t_s^4)\tilde{\Lambda}_k^{-3}-75(-1+t_s^2)^2\tilde{\Lambda}_k^{-2}\nonumber\\&+&\Big\{5-17t_s^2+t_s^4\big\{25+\lvert t_s \rvert(16+\lvert t_s \rvert(85+2\lvert t_s \rvert(8+15\lvert t_s \rvert)))\big\}\Big\}\bigg]\sin\gamma \cos\gamma
\end{eqnarray}
\end{widetext}

\onecolumngrid\
\begin{center}\
\begin{figure}
\begin{tabular}{ccc}
\includegraphics[width=0.5\linewidth]{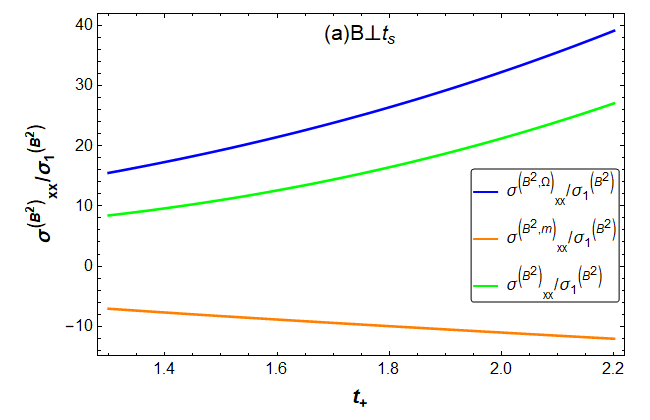} \
\includegraphics[width=0.5\linewidth]{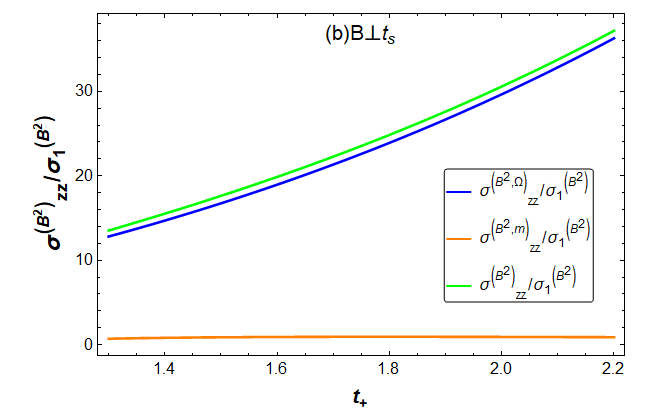} \    
\end{tabular}
\caption{(a) The tilt $t_+$ dependence of the quadratic B conductivity $\sigma_{xx}^{(B^2)}$  .(b) The tilt $t_+$ dependence of the quadratic B conductivity $\sigma_{zz}^{(B^2)}$ for the case of  $\bm B \perp \hat{\bm t}_s$ at B=1 T and $\gamma=\pi/4$ . All the other parameters are identical to those in Fig.(\ref{fig_cond_noB_tilt}).}
\label{cond_xx_quadraticB_prependicular}\
\end{figure}\
\end{center}\
\twocolumngrid\

The magnetic orbital moment contributions of conductivities

\begin{widetext}
\begin{eqnarray}
\sigma_{xx}^{(B^2,m)}(\omega)&=&\frac{\sigma_1^{(B^2)}}{8\lvert t_s \rvert^5}\bigg[\Big\{40\tilde{\Lambda}_k^{-6}-24\tilde{\Lambda}_k^{-5}+15(24+t_s^2)\tilde{\Lambda}_k^{-4}+40(-1+t_s^2)\tilde{\Lambda}_k^{-3}-30(-4+t_s^2+3t_s^4)\tilde{\Lambda}_k^{-2}-8+7t_s^2+30t_s^4\nonumber\\&-&16\lvert t_s \rvert^5-85t_s^6\Big\}\cos^2\gamma + 2\Big\{10\tilde{\Lambda}_k^{-6}+12\tilde{\Lambda}_k^{-5}-15(-6+t_s^2)\tilde{\Lambda}_k^{-4}+20\tilde{\Lambda}_k^{-3}-30(-1+t_s^2)\tilde{\Lambda}_k^{-2}-2+5t_s^2\nonumber\\&-&12\lvert t_s \rvert^5+5t_s^6\Big\}\sin^2\gamma\bigg]\\
\sigma_{zz}^{(B^2,m)}(\omega)&=&\frac{\sigma_1^{(B^2)}}{2\lvert t_s \rvert^5}\bigg[-15\tilde{\Lambda}_k^{-6}+15(-9+t_s^2)\tilde{\Lambda}_k^{-4}+45(-1+t_s^2)\tilde{\Lambda}_k^{-2}+3-8t_s^2+5t_s^4\bigg]\\
\sigma_{xy}^{(B^2,m)}(\omega)&=&\frac{\sigma_1^{(B^2)}}{8\lvert t_s \rvert^5}\bigg[20\tilde{\Lambda}_k^{-6}-48\tilde{\Lambda}_k^{-5}+45(4+t_s^2)\tilde{\Lambda}_k^{-4}+40(-2+t_s^2)\tilde{\Lambda}_k^{-3}+30(2+t_s^2-3t_s^4)\tilde{\Lambda}_k^{-2}-4-3t_s^2+30t_s^4\nonumber\\&+&8\lvert t_s \rvert^5-95t_s^6\bigg]\sin\gamma \cos\gamma
\end{eqnarray}
\end{widetext}

It is noted that all the other magnetoconductivity components are zero. It is clear that the above conductivity equations are independent of chirality, i.e., the Weyl cones with opposite chiralities have the same contribution to the conductivity. Figure(\ref{cond_xx_quadraticB_prependicular}a) demonstrates that the OMM suppresses the longitudinal conductivity over the entire range of tilt parameters. This cancellation reflects the fact that both BC and OMM conductivity components originate from the same underlying Bloch-band geometry but affect the carrier dynamics in opposite ways. Figure~(\ref{cond_xx_quadraticB_prependicular}b) shows that the magnetic-field-induced conductivity component \(\sigma_{zz}^{(B^2)}\) exhibits a behavior distinct from that of \(\sigma_{xx}^{(B^2)}\). In the type-II WSM regime, the overtilted Weyl cones give rise to coexisting electron and hole pockets at the Fermi energy, resulting in an open Fermi surface and a finite density of states at the Weyl node. These features substantially modify the momentum-space integrals entering the conductivity, leading to a positive contribution from the OMM and, consequently, an enhancement of the total conductivity. In contrast, for type-I WSMs, the OMM suppresses both longitudinal conductivity components, \(\sigma_{xx}^{(B^2)}\) and \(\sigma_{zz}^{(B^2)}\), as reported in Ref.~\cite{gao2022suppression}.

In the present system, the quadratic B planar Hall effect can take place in WTe$_2$ at low temperature and manifest itself in a nonzero conductivity $\sigma_{xy}^{(B^2 )} (\omega)$ \cite{das2019linear, burkov2017giant, nandy2017chiral, ma2019planar, kumar2018planar, yang2019current,chen2018planar,li2019anisotropic}. Similar to the diagonal components of the conductivity, $\sigma_{xy}^{(B^2 )}(\omega)$ also consists of the contributions from the BC and the OMM. Figure(\ref{cond_xy_quadraticB_prependicular}a) shows the effect of the tilt on the planar Hall magnetoconductivity. It is seen that the total planar Hall conductivity is suppressed in type-II WSMs when the OMM is present. Figure (\ref{cond_xy_quadraticB_prependicular}b) shows the planar Hall magnetoconductivity as a function of the THz incident light at $\gamma= \pi/4$. We find that $\sigma^{(B^2)}_{xy}$ is smaller in magnitude in the type-II phase and is about four orders of magnitude smaller
than $\sigma^{(0)}$. The real and imaginary parts of total conductivities are suppressed due to the negative contribution of the real and imaginary  parts of OMM conductivities. The suppression predicted here should be experimentally observable through terahertz magneto-optical spectroscopy and planar Hall measurements in tilted type-II WSMs. Since the effect increases with the tilt parameter, strongly tilted type-II materials are expected to exhibit the largest OMM-induced corrections.\\

\subsection{Hall conductivities $\sigma_{ab}^{(H,0)}$ and $\sigma_{ab}^{(H,B)}$}
The Hall response provides one of the most direct manifestations of the topological
character of Weyl semimetals. In contrast to conventional metals, where the Hall effect
originates solely from the Lorentz force acting on charge carriers, Weyl semimetals
possess an additional intrinsic contribution arising from the BC. Consequently,
the Hall conductivity consists of both anomalous and magnetic-field-induced components,
allowing geometric properties of the electronic band structure to be directly probed through
transport measurements.\\

Within the semiclassical Boltzmann formalism, the total Hall conductivity can be expressed
 by the second term of Eq.(\ref{cur_den}),

\begin{equation}
\sigma_{ab}^{(H)}=\sigma_{ab}^{(H,0)}+\sigma_{ab}^{(H,B)}\label{hall_cond}
\end{equation}

where
\begin{eqnarray}
\sigma_{ab}^{(H,0)}(\omega)&=&-\frac{e^2}{\hbar(2\pi)^3}\epsilon_{abc}\int d^3k \Omega_c^s f_0^s\label{Berry_Hall}\\
\sigma_{ab}^{(H,B)}(\omega)&=&\frac{e^2}{\hbar(2\pi)^3}\epsilon_{abc}\int d^3k \Omega_c^s(\textbf{m}_{\bm k}^s\cdot \textbf{B})\Bigl(\frac{\partial f_0^s}{\partial \epsilon_{\bm k}^s}\Bigr)\label{Ordinary_Hall}
\end{eqnarray}

\noindent where $ \epsilon_{abc} $ is the Levi-Civita symbol with $a, b, c,\in  {x, y, z}$
The first term $ \sigma_{ab}^{(H,0)}(\omega) $ in Eq.(\ref{hall_cond}), referring to the local anomalous Hall conductivity associated with the BC, is not equivalent to zero only in the system with broken time-reversal symmetry \cite{xiao2010berry, gao2022suppression}. While the second term $\sigma_{ab}^{(H,B)}(\omega)$ stands for the ordinary Hall conductivity linear in external magnetic field($\bm B$) and describes the leading magnetic-field correction before Landau quantization becomes significant\cite{morimoto2016semiclassical}.\\

This distinction is particularly important for type-II WSMs because the anomalous Hall response is generally non-universal and can depend on tilt and Fermi-surface structure. Zyuzin and Tiwari \cite{zyuzin2016intrinsic} found, using a two-node Kubo treatment, that the anomalous Hall conductivity of type-II Weyl systems is not universal and can depend on the cone tilts; they also showed that finite Hall response can arise from unequal tilts even when the node separation vanishes. Menon and Basu \cite{Menon2020Anomalous} subsequently considered separated multi Weyl nodes and found that the type-II Hall response involves the Fermi-surface structure and an ultraviolet/quantization cutoff in their treatment. Our results are complementary to these works rather than a replacement for them. The present work focuses on semiclassical magneto-optical transport and, in particular, on the OMM correction to the magnetic-field-dependent conductivity.\\

\onecolumngrid\
\begin{center}\
\begin{figure}
\begin{tabular}{ccc}
\includegraphics[width=0.5\linewidth]{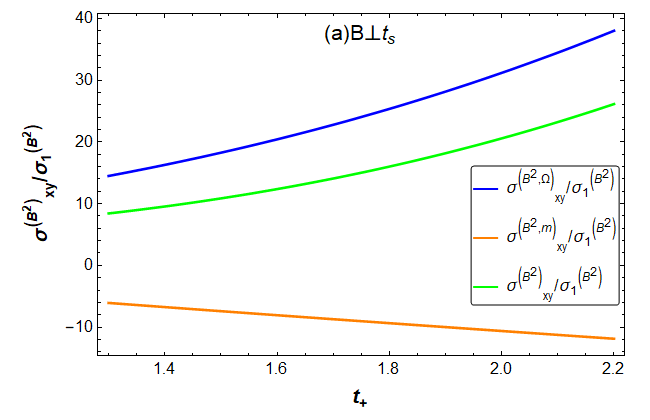} \
\includegraphics[width=0.5\linewidth]{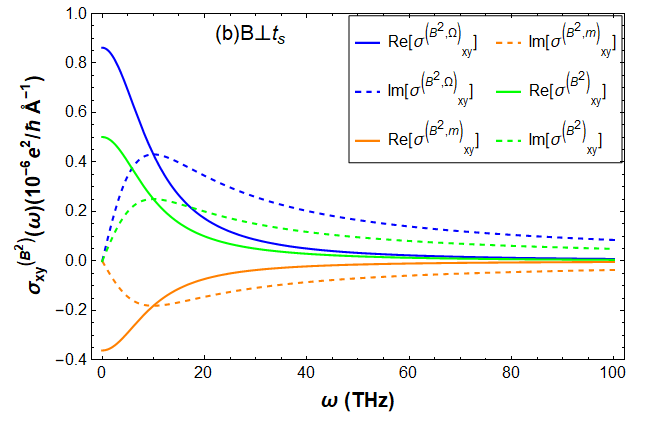} \    
\end{tabular}
\caption{(a) The tilt $t_+$ (b) The frequency dependence of the quadratic B planar Hall conductivity $\sigma_{xy}^{(B^2)}$ at $t_+=1.3$ for the case of  $\bm B \perp \hat{\bm t}_s$ at B=1 T and $\gamma=\pi/4$. All the other parameters are identical to those in Fig.(\ref{fig_cond_noB_tilt}).}
\label{cond_xy_quadraticB_prependicular}\
\end{figure}\
\end{center}\
\twocolumngrid\

Equations (\ref{Berry_Hall}) and (\ref{Ordinary_Hall}) show that the Hall conductivity depends explicitly on both the magnetic-field orientation and the electronic structure of the Weyl cone. To illustrate the angular dependence, the magnetic field is written in spherical coordinates,  $ \textbf{B}=(B_x,B_y,B_z)=(B \sin\theta \cos\phi, B \sin\theta \sin\phi, B\cos\theta ) $ with $\theta=\cos^{-1}(B_z/B)$,  $\phi=\tan^{-1}(B_y/B_x)$. For a single Weyl node, the B-linear contribution to the Hall conductivity can be written as \\

$\sigma^{(H,B)}$\\
\begin{equation}
=
\left(
\begin{array}{ccc}
0 &\sigma_1^H \cos \theta &-\sigma_1^H \sin \theta \sin \phi \\
-\sigma_1^H \cos \theta  & 0& \sigma_1^H \sin \theta \cos \phi \\
\sigma_1^H \sin \theta \sin \phi &-\sigma_1^H \sin \theta \cos \phi&0
\end{array}
\right)   \label{eq1}
\end{equation}%

where

\begin{equation}
\sigma_{1}^{H}=-\frac{e^2}{\hbar}\frac{e B \mathit{v}_F(3\tilde{\Lambda
}_k^{-2}+1)}{24\pi^2  \lvert t_s\rvert^3 \mu}
\end{equation}

\begin{figure}
\begin{tabular}{c}
\includegraphics[width=1\linewidth]{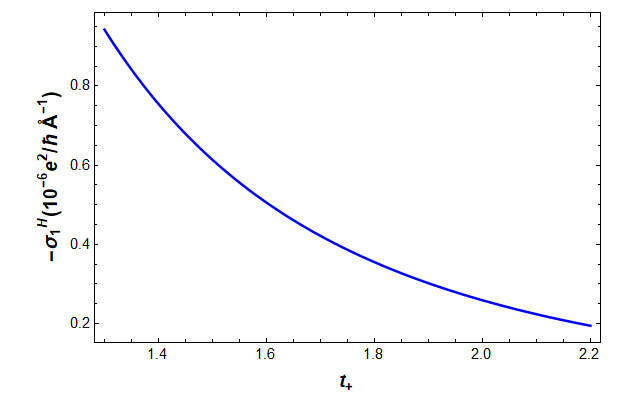} \
\end{tabular}
\caption{The tilt $t_+$  dependence of ordinary Hall conductivity at B=1 T. All the other parameters are identical to those in Fig.(\ref{fig_cond_noB_tilt}).} 
\label{cond_Hall}\
\end{figure}\

Clearly, the linear B contribution to the Hall conductivity of tilted WSMs does not depend on the chirality of the valley and is obtained by summing over all the Weyl nodes. In contrast to the type-I WSM node,  the B-linear contribution to the Hall conductivity of type-II WSMs depends on the tilt parameter $t_s$ as shown in Fig.(\ref{cond_Hall})\cite{gao2022suppression}. The type-II phase displays a pronounced decrease in Hall conductivity as the tilt parameter increases. Physically, this behavior originates from the increasing asymmetry of the electronic states participating in transport. These results establish Hall transport as a sensitive probe of the geometric properties of tilted Weyl fermions and suggest experimentally accessible signatures of the type-II phase.\\

The linear dispersion near a Weyl node gives rise to a monopole-like BC that becomes singular at the nodal point. As a consequence, the corresponding Hall response may exhibit an inverse dependence on the chemical potential, ($\sigma_{xy}\propto 1/\mu$), leading to an apparent divergence as ($\mu\rightarrow 0$). In realistic systems, this divergence is regularized by the finite energy broadening associated with a nonzero relaxation time ($\tau$) which introduces an effective lower energy scale for the Fermi energy \cite{morimoto2016semiclassical}. 
The apparent divergence of the Hall conductivity as the chemical potential approaches the Weyl-node energy should be interpreted with caution. Although a finite relaxation time introduces an energy-broadening scale \(\Gamma\sim\hbar/\tau\), it does not, by itself, guarantee a complete regularization of the divergence. The BC and OMM become singular near an isolated Weyl node, where the validity of the semiclassical approximation may break down. In addition, the semiclassical description requires the electric-field-induced momentum shift during the relaxation time to remain small compared with the characteristic Fermi momentum, \(eE\tau/\hbar\ll k_F\). Therefore, the response sufficiently close to the nodal energy may require a treatment that accounts for disorder broadening and interband effects beyond the semiclassical approximation. This infrared limitation is distinct from the ultraviolet cutoff dependence associated with the open Fermi surfaces of type-II Weyl semimetals. A similar dependence on the chemical potential also emerges in the second-order nonlinear Hall response, as discussed below.\\

It is important to emphasize that the present treatment differs fundamentally from quantum
mechanical descriptions based on Landau-level quantization. The expressions derived above are valid in the weak magnetic fields and small cyclotron frequency $\omega_c$, where the Landau quantization can be ignored. It is valid in the regime $\omega_c \tau \ll\ 1$,  $\hbar\omega_c\ll\mu$($\omega_c$=$e B v_F^2/\mu$) and many Landau levels remain occupied. Under these conditions, the electronic motion is well described by semiclassical wave packets, and the Hall response originates primarily from BC corrections to the carrier dynamics. At sufficiently strong magnetic fields, however, discrete Landau levels dominate the electronic spectrum, leading to quantum oscillations and quantum Hall phenomena that cannot be captured within the present framework. Therefore, the semiclassical theory
should be viewed as complementary to Landau-level approaches: it accurately describes
finite-frequency magneto-optical transport in the weak-field regime, whereas quantum
mechanical methods become essential in the ultra-quantum limit.\\


\section{Second order non-linear response of tilted-WSMs}\label{non-linear}
Now, we explore the second-order nonlinear magneto-optical response of Weyl semimetals. It has emerged as one of the most sensitive probes of the geometric properties of Bloch bands in topological quantum materials. For tilted type-II WSMs, the OMM can either suppress or enhance specific SHG tensor components depending on the field orientation and tilt direction.\\

Substituting  Eq.(\ref{eomdecoupleb}) into Eq.(\ref{SBE}), neglecting the Lorentz force term, and retaining terms up to second order in $\textbf{ E}$, we obtain
\begin{eqnarray}
\frac{1}{\hbar D}[-e \textbf{E}-\frac{e^2}{\hbar}(\textbf{E}\cdot\textbf{B})\bm\Omega_{\bm k}^s]\cdot \frac{\partial \tilde{f}_1^s}{\partial \bm k} -i \omega \tilde{f}_2^s=-\frac{\tilde{f}_2^s}{\tau}
\end{eqnarray}

Solve for $ \tilde{f}_2^s $, we obtain\\

\begin{equation}\label{eqf2}
\tilde{f}_2^s=\frac{\tau}{(1-2i\omega \tau)}\frac{1}{\hbar D}[e \textbf{E}+\frac{e^2}{\hbar}(\textbf{E}\cdot\textbf{B})\bm\Omega_{\bm k}^s]\cdot \frac{\partial \tilde{f}_1^s}{\partial \bm k}
\end{equation}

Substituting Eq.(\ref{disf}) into Eq.(\ref{eqf2}) and retaining terms up to first order in $ \textbf{B} $ yields the second-order nonequilibrium distribution function
\begin{widetext}
\begin{eqnarray}\label{eqnf2}
\tilde{f}_2^s&=&\frac{e\tau^2}{\hbar(1-i\omega \tau)(1-2i\omega \tau)}\biggl\lbrace\textbf{E}\cdot \frac{\partial}{\partial \bm k}\biggr[\Bigl(e \textbf{E}+\frac{e^2}{\hbar}(\textbf{E}\cdot \textbf{B})\bm\Omega_{\bm k}^s-\frac{e^2}{\hbar}(\textbf{B}\cdot \bm\Omega_{\bm k}^s)\textbf{E}\Bigr)\cdot \bm{\mathit{v}}_{\bm k}^s \frac{\partial f_0^s}{\partial \epsilon_{\bm k}^s}-\frac{e}{\hbar}\textbf{E}\cdot\frac{\partial}{\partial \bm k}\biggl(\textbf{m}_{\bm k}^s\cdot \textbf{B}\frac{\partial f_0^s}{\partial \epsilon_{\bm k}^s}\biggr)\biggr]\nonumber\\
&+&\biggl[\frac{e^2}{\hbar}(\textbf{E}\cdot \textbf{B})\bm\Omega_{\bm k}^s -\frac{e^2}{\hbar}(\textbf{B}\cdot \bm\Omega_{\bm k}^s)\textbf{E}\biggr]\cdot\frac{\partial}{\partial \bm k}\biggl(\textbf{E}\cdot \bm{\mathit{v}}_{\bm k}^s \frac{\partial f_0^s}{\partial \epsilon_{\bm k}^s}\biggr)\biggr\rbrace
\end{eqnarray}
\end{widetext}

Now the electric current density at the frequency $ 2\omega $ is given by

\begin{eqnarray}\label{cur_den_2}
\bm{j}_2&=&-\frac{e}{(2\pi)^3}\int d^3k \Bigl[\bm{\mathit{\tilde{v}}}_{\bm k}^s+\frac{e}{\hbar}(\bm\Omega_{\bm k}^s \cdot \bm{\mathit{\tilde{v}}}_{\bm k}^s )\textbf{B}\Bigr]\tilde{f}_2^{s} \nonumber\\
&-&\frac{e^2}{2\pi)^3\hbar}\int d^3k \textbf{E}\times \bm\Omega_{\bm k}^s \tilde{f}_1^s
\end{eqnarray}

According to the definition of second harmonic conductivity, this equation should be written in the form

\begin{equation}
\textbf{j}(2\omega)=\sigma(2\omega)\textbf{E}(\omega)\textbf{E}(\omega)
\end{equation}
where $\sigma(2\omega)  $ is the second harmonic conductivity.\\

The second-order conductivity naturally separates into two physically distinct contributions
\begin{equation}
\sigma_{abc}(2\omega)=\sigma_{abc}^0(2\omega) + \sigma_{abc}^B(2\omega)
\end{equation}
where
$\sigma_{abc}^0(2\omega)$ is the nonlinear conductivity in the absence of a magnetic field and $\sigma_{abc}^B(2\omega)$ represents the magnetic-field-induced correction.

\subsection{Second harmonic conductivity $\sigma_{abc}^{0}$ in the absence of magnetic field}
In this subsection, we calculate the second harmonic current of the Weyl semimetal system in the absence of magnetic field \textbf{B}. Inserting Eq. (\ref{eqnf2}) into the first term of Eq. (\ref{cur_den_2}), the second harmonic conductivity tensor can be written as

\begin{equation}
\sigma_{abc}^0(2\omega)
=\frac{-\tau^{2}}{(1-i\omega \tau)(1-2i\omega \tau)}\frac{e^3}{\hbar(2\pi)^3}\int d^3k \frac{\partial \mathit{v}_a^s}{\partial k_c}  \mathit{v}_b^s
\Bigl(-\frac{\partial f_0^s}{\partial \epsilon_{\bm k}^s}\Bigr)
\end{equation}

The symmetry of the tilted Weyl Hamiltonian allows only a limited number of independent
tensor components. Specifically, the aaz, aza, and zaa (a = x, y, z) components of the second harmonic conductivity tensor are nonzero and all other components equal to zero \cite{gao2022suppression,gao2021second}. This selection rule follows directly from the rotational symmetry around the tilt axis and the absence of inversion symmetry required for second-order optical processes.\\

\begin{widetext}
\begin{eqnarray}
\sigma_{zzz}^0(2\omega)&=&\frac{2\sigma_{DL}}{t_s^4}\sgn(t_s) \bigg[-\tilde{\Lambda}_k^{-2}+(-3+t_s^2)-3(-1+t_s^2)\ln[(t_s^2-1)\tilde{\Lambda}_k^2]\bigg]\\
\sigma_{xxz}^0(2\omega)&=&\frac{ \sigma_{DL}}{t_s^4}\sgn(t_s) \bigg[\tilde{\Lambda}_k^{-2}-(-3+t_s^2)+(-3+t_s^2)\ln[(t_s^2-1)\tilde{\Lambda}_k^{2}]\bigg]
\end{eqnarray}
\end{widetext}

where
\begin{equation}
\sigma_{DL}=\frac{e^3\tau^2\mu}{(1-2i\omega \tau)(1-i\omega \tau)8\pi^2\hbar^3}
\end{equation}
is the Drude-like frequency dependent complex conductivity.\\

\onecolumngrid\
\begin{center}\
\begin{figure}
\begin{tabular}{ccc}
\includegraphics[width=0.5\linewidth]{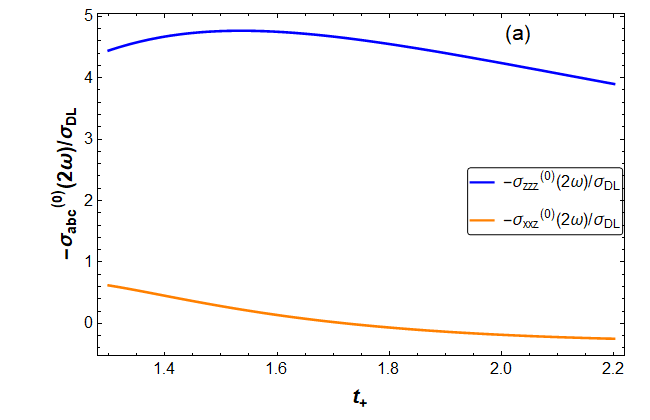} \
\includegraphics[width=0.5\linewidth]{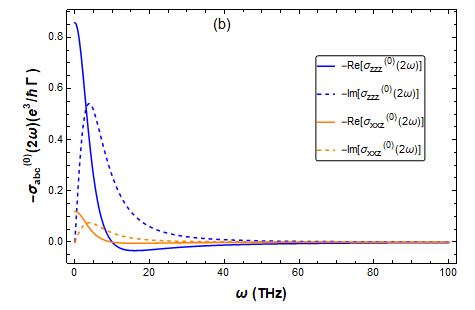} \    
\end{tabular}
\caption{(a) The tilt  $t_+$ dependence of the nonlinear conductivities $\sigma_{zzz}^0$ and $\sigma_{xxz}^0$.(b) The frequency dependence of conductivities $\sigma_{zzz}^0$ and $\sigma_{xxz}^0$  for the process of second harmonic generation at $t_+=1.3$. Here the relaxation rate $\Gamma=\hbar/\tau$. All the other parameters are identical to those in Fig.(\ref{fig_cond_noB_tilt}).}
\label{cond_noB_second_harmonic_tilt}\
\end{figure}\
\end{center}\
\twocolumngrid\

The second-harmonic conductivity tensor obeys the symmetry relations $\sigma_{zxx}^0(2\omega)
=\sigma_{zyy}^0(2\omega)=\sigma_{xzx}^0(2\omega)=\sigma_{yzy}^0(2\omega)=\sigma_{xxz}^0(2\omega)=\sigma_{yyz}^0(2\omega)$. Moreover, these conductivity components are independent of the chirality of an individual Weyl node. The total second-harmonic response of a tilted Weyl semimetal is obtained by summing the contributions from a pair of Weyl nodes. In the presence of equal tilt configuration  $t_{+} = -t_{-}$ the contributions from the two nodes cancel, yielding $\sigma_{abc}(2\omega)= 0 $. By contrast, for the opposite tilt configuration, $t_{+} = t_{-}$, the node contributions add constructively, resulting in a finite second-harmonic conductivity, $\sigma_{abc}(2\omega)\neq 0$.\\

As shown in Ref.\cite{gao2022suppression}, the second-harmonic conductivity $\sigma_{abc}^{0}(2\omega)$ vanishes identically in the untilted limit, $t_s\rightarrow 0$, demonstrating that a finite tilt of the Weyl cone is essential for generating a second-harmonic response. The tilt modifies the Fermi-surface geometry and, for type-I Weyl semimetals, transforms the point-like Fermi surface at the nodal energy into an ellipsoidal one at finite chemical potential. In type-II Weyl semimetals, the overtilted dispersion instead produces an open, hyperbolic Fermi surface within the linearized model. These distinct Fermi-surface geometries are directly related to the contrasting transport and nonlinear optical responses of the two phases.\\

It is further evident from Fig.(\ref{cond_noB_second_harmonic_tilt}a) that $\sigma_{zzz}^{0}(2\omega)$ exhibits a stronger sensitivity to the tilt parameter than $\sigma_{xxz}^{0}(2\omega)$. The conductivity components $\sigma_{zzz}^{0}(2\omega)$ exhibit nonmonotonic tilt dependences while $\sigma_{xxz}^{0}(2\omega)$ has suppressed behavior. Finally, Fig.(\ref{cond_noB_second_harmonic_tilt}b) depicts the frequency dependence of the second-harmonic conductivity. Different from the linear response regime, the real parts of $\sigma^{(0)}_{abc}$ exhibit positive and negative amplitudes because of the complex denominator function $\sigma_{DL}$ contribution to the second harmonic
generation processes.

\subsection{Second harmonic conductivity $\sigma_{abc}^{B}$ in the presence  of linear magnetic field}
Inserting Eq. (\ref{eqnf2}) into the first term of Eq. (\ref{cur_den_2}), we write the linear $ \textbf{B} $ dependent second harmonic conductivity tensor 
\begin{widetext}
\begin{eqnarray}\label{second_B_cond}
\sigma_{abc}^{(B)}(2\omega)&=&\frac{\tau^{2}}{(1-i\omega \tau)(1-2i\omega \tau)}\frac{e^3}{\hbar(2\pi)^3}\int d^3k \biggl\lbrace \frac{\partial \mathit{v}_a^s}{\partial k_c}\Bigl[-\frac{e B_b}{\hbar}(\bm\Omega_{\bm k}^s \cdot \bm{\mathit{v}}_{\bm k}^s )+\frac{e \mathit{v}_b^s}{\hbar}(\bm\Omega_{\bm k}^s \cdot \bm{B})\Bigr]-\frac{e B_c}{\hbar}\frac{\partial}{\partial \bm k}\cdot(\mathit{v}_a^s \bm\Omega_{\bm k}^s )\mathit{v}_b^s
\nonumber\\
&+&\frac{e}{\hbar} \frac{\partial [\mathit{v}_a^s(\bm\Omega_{\bm k}^s \cdot \bm{B})]}{\partial k_c}\mathit{v}_b^s-\frac{e B_a}{\hbar}\frac{\partial (\bm\Omega_{\bm k}^s \cdot   \bm{\mathit{v}}_{\bm k}^s )}{\partial k_c}\mathit{v}_b^s +\frac{\partial^2(\textbf{m}_{\bm k}^s\cdot \textbf{B})}{\hbar  \partial k_a \partial k_c}\mathit{v}_b^s -\frac{\partial^2 \mathit{v}_a^s}{\hbar  \partial k_a \partial k_c} (\textbf{m}_{\bm k}^s\cdot \textbf{B})\biggr\rbrace \Bigl(-\frac{\partial f_0^s}{\partial \epsilon_{\bm k}^s}\Bigr)
\end{eqnarray}
\end{widetext}

We will explore Eq.(\ref{second_B_cond}) with following two cases:\\

Case-I In the case of  $\textbf{B}\Vert \hat{\bm t}_s \Vert \hat{\bm z} $, for a single Weyl node, we obtain
the magnetoconductivity components

\begin{widetext}
\begin{eqnarray}
\sigma_{xzx}^{(B)}(2\omega)&=& s\frac{ \sigma_2^{(B)}}{15t_s^5}\bigg[21\tilde{\Lambda}_k^{-5}+15(7-2t_s^2)\tilde{\Lambda}_k^{-4}+15(14-5t_s^2)\tilde{\Lambda}_k^{-3}+30(7-8t_s^2+t_s^4)\tilde{\Lambda}_k^{-2}+15(7-7t_s^2+2t_s^4)\tilde{\Lambda}_k^{-1}\nonumber\\ &+& 21-50t_s^2+45t_s^4-6t_s^6\bigg]
\end{eqnarray}

\begin{eqnarray}
\sigma_{zxx}^{(B)}(2\omega)&=&s\frac{ \sigma_2^{(B)}}{240}\bigg[176-\frac{210}{\lvert t_s \rvert}-80\bigg(1-\frac{(\tilde{\Lambda}_k^{-1}-1)^3}{\lvert t_s \rvert^3}\bigg)-\frac{210}{\lvert t_s \rvert}\tilde{\Lambda}_k^{-1}+55\cos\Big(3\csc^{-1}\Big(\frac{\lvert t_s \rvert}{1+\tilde{\Lambda}_k^{-1}}\Big)\Big)\nonumber\\&-&21\cos\Big(5\csc^{-1}\Big(\frac{\lvert t_s \rvert}{1+\tilde{\Lambda}_k^{-1}}\Big)\Big)\bigg]
\end{eqnarray}
\end{widetext}

\noindent where
\begin{equation}
\sigma_{2}^{(B)}=\frac{e^4\tau^2 B v_F^2}{\hbar^2(1-2i\omega \tau)(1-i\omega\tau)8\pi^2\mu}
\end{equation}

We can check that other nonzero second harmonic conductivity components satisfy the relations: $ \sigma_{zxx}^{(B)}(2\omega)=\sigma_{zyy}^{(B)}(2\omega)=\sigma_{xzx}^{(B)}(2\omega)=\sigma_{yzy}^B(2\omega) $.\\

Case-II In the case of $\textbf{B}\perp \hat{\bm t}_s (\Vert \hat{\bm z}) $, we get the nonzero conductivity
components

\begin{widetext}
\begin{eqnarray}
\sigma_{zxz}^{(B)}(2\omega)&=&s\frac{ \sigma_2^{(B)}}{48\lvert t_s \rvert^3}\bigg[8\Big\{3\tilde{\Lambda}_k^{-4}-17\tilde{\Lambda}_k^{-3}+3(5-2t_s^2)\tilde{\Lambda}_k^{-2}+3(-5+4t_s^2)\tilde{\Lambda}_k^{-1}+\Big\{2+t_s^2(-6+\lvert t_s \rvert(4+3\lvert t_s \rvert))\Big\}\Big\}\nonumber\\&+&\Big\{11\cos\Big[3\sec^{-1}\Big(\frac{\lvert t_s \rvert}{1+\tilde{\Lambda}_k^{-1}}\Big)\Big]-3\cos\Big[5\csc^{-1}\Big(\frac{\lvert t_s \rvert}{1+\tilde{\Lambda}_k^{-1}}\Big)\Big]\Big\}\bigg]\cos\gamma\\
\sigma_{xzz}^{(B)}(2\omega)&=&s\frac{ \sigma_2^{(B)}}{6\lvert t_s \rvert^3}\bigg[9\tilde{\Lambda}_k^{-4}+16\tilde{\Lambda}_k^{-3}+6(9-7t_s^2)\tilde{\Lambda}_k^{-2}-12(-1+t_s^2)\tilde{\Lambda}_k^{-1}+9-18t_s^2-4\lvert t_s \rvert^3+9t_s^4\bigg]\cos\gamma\\
\sigma_{zyz}^{(B)}(2\omega)&=&s\frac{ \sigma_2^{(B)}}{48t_s^3}\bigg[8\Big\{3\tilde{\Lambda}_k^{-4}-17\tilde{\Lambda}_k^{-3}+3(5-2t_s^2)\tilde{\Lambda}_k^{-2}+3(-5+4t_s^2)\tilde{\Lambda}_k^{-1}+\Big\{2+t_s^2(-6+\lvert t_s \rvert(4+3\lvert t_s \rvert))\Big\}\Big\}\nonumber\\&+&\Big\{11\cos\Big[3\sec^{-1}\Big(\frac{\lvert t_s \rvert}{1+\tilde{\Lambda}_k^{-1}}\Big)\Big]-3\cos\Big[5\csc^{-1}\Big(\frac{\lvert t_s \rvert}{1+\tilde{\Lambda}_k^{-1}}\Big)\Big]\Big\}\bigg]\sin\gamma\\
\sigma_{yzz}^{(B)}(2\omega)&=&s\frac{ \sigma_2^{(B)}}{6\lvert t_s \rvert^3}\bigg[9\tilde{\Lambda}_k^{-4}+16\tilde{\Lambda}_k^{-3}+6(9-7t_s^2)\tilde{\Lambda}_k^{-2}-12(-1+t_s^2)\tilde{\Lambda}_k^{-1}+9-18t_s^2-4\lvert t_s \rvert^3+9t_s^4\bigg]\sin\gamma
\end{eqnarray}
\end{widetext}

In both situation, unlike type-I WSMs, the B-linear contribution to the second harmonic conductivity in type-II WSMs is dependent on the tilt. For the opposite tilt configuration $t_{+} = -t_{-}$, the contributions from the two nodes with opposite chirality cancels each other, yielding $\sigma_{abc}(2\omega)= 0 $. By contrast, for an equal tilt configuration, $t_{+} = t_{-}$, the nodes contribution add constructively, resulting in a finite second-harmonic conductivity.\

\subsection{The second order nonlinear Hall conductivity $\sigma_{abc}^{(H,0)}$ in the absence of magnetic field}
In this subsection, we study the second-order nonlinear Hall effect of WSMs without a magnetic field. Inserting Eq. (\ref{disf}) into the second term in Eq. (\ref{cur_den_2}) and taking B = 0, we obtain the nonlinear Hall conductivity.

\begin{equation}
\sigma_{abc}^{(H,0)}=\epsilon_{acd}\frac{e^3 \tau}{\hbar^2(1-i\omega\tau)}D_{bd}
\end{equation}

\noindent where $ D_{bd} $ is called Berry curvature dipole(BCD). It represents the first-order momentum derivative of the BC, which arises from the uneven distribution of BC in momentum space\cite{sodemann2015quantum}. Unlike the conventional linear Hall effect, which relies on an external magnetic field, the BCD facilitates a transverse Hall voltage driven exclusively by an applied electric field. This effect is particularly pronounced in type-II Weyl semimetals due to the strong tilting of the Weyl cones.\\

The BC itself remains unchanged in going from type-I to type-II Weyl semimetals because the tilt term does not alter the Bloch eigenstates. However, the BCD is a Fermi-surface property, not a purely topological invariant. In type-I systems, the $1/k^2$ BC scaling and the spherical Fermi surface lead to an exact cancellation of the Fermi-momentum (and hence $\mu$) dependence within the linear model. In type-II systems, the overtilt destroys this scale invariance by creating anisotropic, open electron and hole pockets. Consequently, the BCD generally depends on the chemical potential through the evolving Fermi-surface geometry, even though the underlying BC retains its monopole form.\\

One can calculate the non-zero components of BCD 

\begin{equation}
D_{bd}=\frac{\hbar}{(2\pi)^3}\int d^3k \Omega_{d}^{s}\mathit{v}_{b}^{s}\Bigl(-\frac{\partial f_0^s}{\partial \epsilon_{\bm k}^s}\Bigr)
\end{equation}

\begin{widetext}
\begin{eqnarray}
D_{xx}=D_{yy}&=&s\frac{1}{16\pi^2\lvert t_s \rvert^3}\bigg[4\tilde{\Lambda}_k^{-1}-2t_s+(-1+t_s^2)\ln\Big(\frac{\lvert t_s \rvert -1}{\lvert t_s \rvert +1}\Big)\bigg]\\
D_{zz}&=&s\frac{1}{8\pi^2\lvert t_s \rvert^3}\biggl[\tilde{\Lambda}_k^{-2}-3(-1+t_s^2)+\ln\tilde{\Lambda}_k^{-1}-\ln[-1+\lvert t_s \rvert]+(-1+t_s^2)\ln[1+\lvert t_s \rvert]\nonumber\\&-&(-1+t_s^2)\ln\tilde{\Lambda}_k^{-1}+t_s^2\ln[(-1+\lvert t_s \rvert)\tilde{\Lambda}_k]\bigg]
\end{eqnarray}
\end{widetext}

In contrast to type-I WSMs, the magnitude of BCD components in type-II WSMs depends on the chemical potential through the momentum cutoff. These components depend on the chirality of the Weyl nodes. For the opposite-tilt configuration, \(t_+=-t_-\), the contributions from the two Weyl nodes with opposite chirality cancel, resulting in a vanishing total Hall conductivity within the present two-node model. In contrast, for the equal-tilt configuration, \(t_+=t_-\), the node-resolved contributions add constructively, producing a finite conductivity\cite{gao2022suppression, sodemann2015quantum, gao2021second}.\\

\subsection{Linear B-contribution to the second order nonlinear Hall conductivity $\sigma_{abc}^{(H,B)}$}
Now, we focus on the second-order nonlinear Hall effect in a weak magnetic field. Inserting Eq.(\ref{disf}) into the second term of Eq.(\ref{cur_den_2}) and retaining terms up to the first power of B, one obtains the complex nonlinear Hall conductivity 

\begin{equation}
\sigma_{abc}^{(H,B)}=\epsilon_{acd}\frac{e^3 \tau}{\hbar^2(1-i\omega\tau)}\bigl[D_{bd}^{\Omega}+D_{bd}^{m}\bigr]\label{cond_berrydip}
\end{equation}

where

\begin{equation}
D_{bd}^{\Omega}=\frac{e}{(2\pi)^3}\int d^3k \Omega_{d}^{s}[B_b(\bm\Omega_{\bm k}^s \cdot   \bm{\mathit{v}}_{\bm k}^s )-\mathit{v}_{b}^{s}(\bm\Omega_{\bm k}^s \cdot \bm{B})]\Bigl(-\frac{\partial f_0^s}{\partial \epsilon_{\bm k}^s}\Bigr)
\end{equation}

\begin{equation}
D_{bd}^{m}=\frac{1}{(2\pi)^3}\int d^3k \frac{\partial \Omega_{d}^{s}}{\partial k_{b}}(\bm m_{\bm k}^s \cdot \bm{B})\Bigl(-\frac{\partial f_0^s}{\partial \epsilon_{\bm k}^s}\Bigr)
\end{equation}

\noindent where $D_{bd}^{\Omega}$ and $D_{bd}^{m}$ are BCD contributions due to the BC and the OMM respectively.\\

\begin{figure}[htbp]
\includegraphics[scale=0.52]{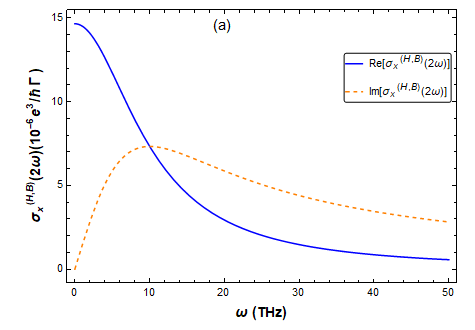} 
\end{figure}
\begin{figure}[htbp]
\includegraphics[scale=0.40]{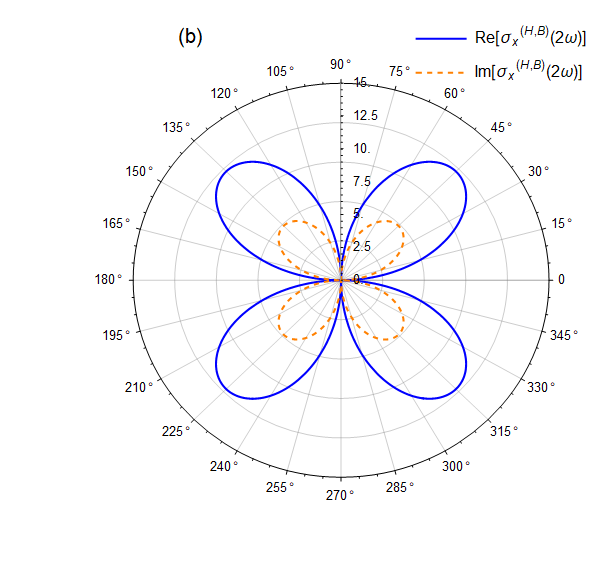} \
\caption{(a)The nonlinear Hall conductivities $\sigma_{x}^{(H,B)}$ for the process of second harmonic generation as a function of the incident photon frequency on the tilt at $t_+=1.3$, B=1 T and $\gamma=\pi/4$.(b) The angle dependence of the nonlinear Hall conductivity $\sigma_{x}^{(H,B)}$ at $\omega=5$ THz. Here the relaxation rate $\Gamma=\hbar/\tau$. All the other parameters are identical to those in Fig.(\ref{fig_cond_noB_tilt}).}
\label{cond_nonlinear_hall_omega}\
\end{figure}\
Again, we will consider the following two cases.\\

Case-I In the case of $\textbf{B}\Vert \hat{\bm t}_s \Vert \hat{\bm z} $, one obtains the BCD components.

\begin{eqnarray}
D_{xx}^{\Omega}=D_{yy}^{\Omega}&=&-\frac{D_2^{(B)}}{2t_s^4}\sgn(t_s) \Big[3\tilde{\Lambda}_k^{-5}-5(-3+t_s^2)\tilde{\Lambda}_k^{-3}+2\lvert t_s \rvert^5\Big]
\nonumber\\
\end{eqnarray}
\begin{eqnarray}
D_{zz}^{\Omega}&=&\frac{D_2^{(B)}}{t_s^4}\sgn(t_s) \Big[3\tilde{\Lambda}_k^{-5}-5(-3+t_s^2)\tilde{\Lambda}_k^{-3}+2\lvert t_s \rvert^5\Big]\nonumber\\
\end{eqnarray}
\begin{eqnarray}
D_{xx}^{m}=D_{yy}^{m}&=&\frac{D_2^{(B)}}{8t_s^4}\sgn(t_s) \Big[135\tilde{\Lambda}_k^{-4}-30(-3+t_s^2)\tilde{\Lambda}_k^{-2}\nonumber\\&-&9+10t_s^2+15t_s^4\Big]\nonumber\\
\end{eqnarray}
\begin{eqnarray}
D_{zz}^{m}&=&-\frac{D_2^{(B)}}{4t_s^4}\sgn(t_s) \Big[135\tilde{\Lambda}_k^{-4}-30(-3+t_s^2)\tilde{\Lambda}_k^{-2}-9+10t_s^2\nonumber\\&+&15t_s^4\Big]
\end{eqnarray}

\noindent where 
\begin{equation}
D_2^{(B)}=\frac{1}{8\pi^2}\frac{e B }{\hbar}\frac{\hbar^2\mathit{v}_F^2}{15\mu^2}
\end{equation}
  
Case-II In the case of $\textbf{B}\perp \hat{\bm t}_s (\Vert \hat{\bm z}) $, we get the nonzero components

\begin{widetext}
\begin{eqnarray}
D_{zx}^{\Omega}&=&-\frac{3D_2^{(B)}}{8t_s^4}\sgn(t_s) \Big[5(-3+t_s^2)\tilde{\Lambda}_k^{-4}-10(-1+t_s^2)^2\tilde{\Lambda}_k^{-2}+\Big\{1+5t_s^2(-1+3t_s^2+t_s^4)\Big\}\Big]\cos\gamma\\
D_{xz}^{\Omega}&=&\frac{D_2^{(B)}}{4t_s^4}\sgn(t_s) \Big[6\tilde{\Lambda}_k^{-5}-15(-3+t_s^2)\tilde{\Lambda}_k^{-4}-10(-3+t_s^2)\tilde{\Lambda}_k^{-3}-30(-1+t_s^2)\tilde{\Lambda}_k^{-2}\nonumber\\&+&\big\{-3+5t_s^2+t_s^4(15+\lvert t_s \rvert(4+15\lvert t_s \rvert))\big\}\Big]\cos\gamma\\
D_{zy}^{\Omega}&=&-\frac{ 3D_2^{(B)}}{8t_s^4}\sgn(t_s) \Big[5(-3+t_s^2)\tilde{\Lambda}_k^{-4}-10(-1+t_s^2)^2\tilde{\Lambda}_k^{-2}+\big\{1+5t_s^2(-1+3t_s^2+t_s^4)\big\}\Big]\sin\gamma\\
D_{yz}^{\Omega}&=&\frac{D_2^{(B)}}{4t_s^4}\sgn(t_s) \Big[6\tilde{\Lambda}_k^{-5}-15(-3+t_s^2)\tilde{\Lambda}_k^{-4}-10(-3+t_s^2)\tilde{\Lambda}_k^{-3}-30(-1+t_s^2)\tilde{\Lambda}_k^{-2}\nonumber\\&+&\big\{-3+5t_s^2+t_s^4(15+\lvert t_s \rvert(4+15\lvert t_s \rvert))\big\}\Big]\sin\gamma
\end{eqnarray}
\end{widetext}
and
\begin{widetext}
\begin{eqnarray}
D_{zx}^{m}&=&\frac{3 D_2^{(B)}}{8t_s^4}\sgn(t_s)\Big[45\tilde{\Lambda}_k^{-4}-30(-1+t_s^2)\tilde{\Lambda}_k^{-2}-3+10t_s^2-15t_s^4\Big]\cos\gamma\nonumber\\
D_{zy}^{m}&=&\frac{3 D_2^{(B)}}{8t_s^4}\sgn(t_s)\Big[45\tilde{\Lambda}_k^{-4}-30(-1+t_s^2)\tilde{\Lambda}_k^{-2}-3+10t_s^2-15t_s^4\Big]\sin\gamma\nonumber\\
\end{eqnarray}
\end{widetext}

\noindent One can check that $D_{zx}^{m} =D_{xz}^{m} $ and $D_{zy}^{m}=D_{yz}^{m}  $ and rest of the components will vanish.\\

The nonlinear Hall effect can be modulated by the polarization of the incident light. Using Eq. (\ref{cond_berrydip}), the electric current is rewritten in the form of

\begin{equation}\label{current_density}
\textbf{j}(2\omega)=\frac{e^3\tau}{\hbar^2(1-i\omega \tau)}(\hat{\textbf{D}}\cdot \textbf{E})\times \textbf{E}
\end{equation}

Assume that an electromagnetic wave propagates in the x direction:
\begin{equation}\label{e_field}
\textbf{E}(\textbf{r},t)=\vert\textbf{E}(\omega)\vert Re[\vert \psi\rangle e^{i(q x- \omega t)}
\end{equation}

where
\begin{equation}
\vert \psi\rangle=\Bigl(
\begin{array}{c}
\psi_y\\
\psi_z
\end{array}\Bigr)=\Bigl(
\begin{array}{c}
\sin \theta e^{i \alpha_y}\\
\cos \theta e^{i \alpha_z}
\end{array}\Bigr)
\end{equation}

\noindent is the Jones vector in the y-z plane with phases $ \alpha_y ,\alpha_z$ and
the amplitudes $E_y = \vert \textbf{E}\vert \sin \theta$ and $E_z = \vert \textbf{E}\vert \cos \theta$. Inserting Eq. (\ref{e_field}) into Eq. (\ref{current_density}), we obtain the nonlinear Hall current

\begin{equation}\label{current_x}
j_{x}(2\omega)=\frac{e^3\tau}{\hbar^2(1-i\omega \tau)}\frac{D_{yy}^{(B)}-D_{zz}^{(B)}}{2}\sin 2\theta e^{(\alpha_y+\alpha_z)}\vert\textbf{E}\vert^{2}
\end{equation}

Eq.(\ref{current_x}) can be rewritten as $ j_x \sim (D_{yy}-D_{zz})E_y E_z$. Hence, conductivity will be

\begin{widetext}
\begin{eqnarray}\label{sigma_B_typeII}
\sigma_{x}^{(H,B)}(2\omega)&=&\frac{3e^3\tau}{16t_s^4(1-i\omega\tau)\hbar^2}\sgn(t_s)\Big[-12\tilde{\Lambda}_k^{-5}+135\tilde{\Lambda}_k^{-4}+20(-3+t_s^2)\tilde{\Lambda}_k^{-3}-30(-3+t_s^2)\tilde{\Lambda}_k^{-2}-9+10t_s^2+15t_s^4\nonumber\\&-&8\lvert t_s \rvert^5\Big] D_2^{(B)}\sin2\theta
\end{eqnarray}
\end{widetext}

Fig.(\ref{cond_nonlinear_hall_omega}a) shows the variation of linear-B second harmonic conductivities $\sigma_{x}^{(H,B)}(2\omega)$ as function of incident photon frequency $\omega$. In contrast to the case of $B = 0$, the B-linear contribution to the nonlinear Hall conductivity is independent of the chirality and the odd function of $t_s$. Only in the system with equal tilt configuration ($t_{+} = t_{-} $ ), the conductivity $\sigma_{x}^{(H,B)}(2\omega) \neq 0$. From Eq.(\ref{sigma_B_typeII}), evidently, $\sigma_{x}^{(H,B)}(2\omega)$  reaches its maximum when the polarization direction $\theta= \pm \pi/4$ and vanishes at $\theta = 0, \pi/2$ as further reflected in Fig(\ref{cond_nonlinear_hall_omega}b).\\

Recent experiments have measured the non-linear Hall effect in type-II WSM WTe$_2$ due to its broken inversion symmetry and strong BCD \cite{ma2019observation, du2021nonlinear, shvetsov2019nonlinear}. Another type-II WSM, MoTe$_2$, displays a significant nonlinear Hall effect \cite{zhang2018berry}. The strong tilting of the Weyl points produces a  ``hot spot'' in the BC, which enhances the BCD.\\


\section{Conclusion}\label{conclusion}
In this work, we have presented the linear and nonlinear magnetoconductivity of tilted type-II Weyl semimetals within the semiclassical Boltzmann framework by incorporating both the Berry curvature (BC) and the orbital magnetic moment (OMM) corrections and compared with the existing theoretical work of type-I phase. In the linear response regime, we derive analytical expressions for the OMM-induced magnetoconductivity while in the nonlinear regime we obtain analytical expressions that capture the combined contributions of the BC and the OMM for different magnetic-field configurations. The competition between these two geometric effects governs the magnitude and sign of the transport coefficients and leads to pronounced differences between type-I and type-II Weyl semimetals. The coexistence of electron and hole pockets in the overtilted Weyl cone enlarges the available phase space for electronic transport, thereby amplifying the OMM-induced corrections to both linear and nonlinear conductivity. Consequently, several conductivity components exhibit considerably larger enhancement or suppression than in conventional type-I Weyl semimetals, providing clear transport signatures of the overtilted electronic structure.\\

We further demonstrated that conductivity components in the type-II case exhibit explicit dependence on the ultraviolet momentum cutoff within the continuum model. We have clarified that this cutoff dependence originates from the unbounded electron and hole pockets of the low-energy type-II Weyl Hamiltonian and should therefore be regarded as a consequence of the continuum approximation rather than a universal physical property. In realistic lattice models, the finite Brillouin zone naturally regularizes these contributions while preserving the qualitative transport trends predicted here.\\

The theoretical predictions presented here are directly relevant to recently discovered
type-II Weyl semimetals such as WTe$_2$, MoTe$_2$ and TaIrTe$_4$ where strong cone tilting and large BC effects have already been experimentally established. The predicted modifications of the linear magneto-optical conductivity, planar Hall response and nonlinear optical conductivity should be accessible through present-day terahertz spectroscopy and angular-dependent magnetotransport experiments performed under moderate magnetic fields.\\


Finally, we emphasize that the present analysis has been carried out within the weak-field semiclassical regime where Landau quantization is negligible and electronic transport is well described by wave-packet dynamics. Extending the present theory to the quantum limit where discrete Landau levels dominate the optical response represents an important direction for future work. Additional extensions may include the effects of finite temperature, disorder beyond the constant relaxation-time approximation, intervalley scattering, electron-electron interactions, strain-induced pseudo-magnetic fields and realistic lattice Hamiltonians that naturally eliminate the ultraviolet cutoff.\\

In summary, this work demonstrates that the OMM is an indispensable ingredient in understanding magneto-optical transport in tilted type-II Weyl semimetals. The combined action of the BC and the OMM produces distinctive linear and nonlinear transport signatures in type-II WSMs that are absent in type-I systems. We expect that these findings will stimulate further theoretical and experimental studies of geometric transport phenomena in topological quantum materials and contribute to the development of future topological optoelectronic devices.

\begin{appendix}
\section{DETAILS OF THE CALCULATIONS USING SPHERICAL POLAR COORDINATES}
In this paper, we have focused on the n-doped tilted Weyl semimetals with a positive chemical potential $\mu$. In general, one can decompose the momentum k using spherical co-ordinates as follows
\begin{eqnarray}
k_x&=& k \sin \theta \cos \phi \\
k_y&=& k \sin \theta \sin \phi \\
k_z&=& k\cos \theta
\end{eqnarray}

The general expression for conductivities is written as
\begin{eqnarray}
\sigma &=& \frac{e^2 \tau}{(2\pi)^3}\int_0^{\infty}k^2 dk \int_0^{2\pi}d\phi_{\bm k}\int_0^{\pi} d\theta_{\bm k}\sin \theta_{\bm k} f(k,\theta_{\bm k}, \phi_{\bm k})\nonumber\\
&&\delta(\mu- \hbar[t_s k \cos \theta_{\bm k} \pm v_F k])
\end{eqnarray}

The Jacobian of the transformation is $\mathcal{J} =  k^2 \sin \theta_{\bm k} $, which has been used for analytical expressions of conductivity elements of tilted-WSMs. Substituting $\cos\theta_{\bm k} \longrightarrow x$, the above expression can be rewritten as \\

\begin{eqnarray}
\sigma &=& \frac{e^2 \tau}{(2\pi)^3}\int_0^{\infty}k^2 dk \int_0^{2\pi}d\phi_{\bm k}\int_{-1}^{1} dx f(k,x, \phi_{\bm k})\nonumber\\
&\times &\delta(\mu- \hbar[t_s k x \pm v_F k])
\end{eqnarray}

For type-II WSM, we introduce a finite momentum cut-off $\Lambda_k$ such that $\Lambda_k(1+t_s)>k_F$ in the radial direction k. For the conduction band with $t_s>1$ we obtain the integration limit to be
\begin{equation}
    \bigg(\frac{k_F}{\Lambda_k}-1\bigg)\frac{1}{t_s}\leq{x}\leq{1},
\end{equation}
with $\Lambda_k(t_s+1)>k_F$. The left-hand side term of this inequality is negative for $t_s>0$. When $t_s<-1$, we get
\begin{equation}
{-1} \leq {x} \leq \bigg(\frac{k_F}{\Lambda_k}-1\bigg)\frac{1}{t_s},
\end{equation}
with $\Lambda_k(-t_s+1)>k_F$.
Similarly, we get the following limit for the valence band
\begin{equation}
     \bigg(\frac{k_F}{\Lambda_k}+1\bigg)\frac{1}{t_s}\leq{x}\leq{1} ,
\end{equation}
with $\Lambda_k(t_s-1)>k_F$ for $t_s>1$. For $t_s<-1$, we get the following limit
\begin{equation}
{-1} \leq {x} \leq \bigg(\frac{k_F}{\Lambda_k}+1\bigg)\frac{1}{t_s} ,
\end{equation}
with $-\Lambda_k(t_s+1)>k_F$. For the calculation of conductivities, all these limits are essential for performing all the integrals.

\end{appendix}

\bibliography{tilted_II_WSMs}
\end{document}